\documentclass[aps,pra,twocolumn,showpacs,superscriptaddress,amsmath,amssymb,floatfix]{revtex4}

\usepackage{graphicx} 
\usepackage[a4paper,left=1.5 cm,right=1cm,top=2 cm,bottom=2cm]{geometry}
\usepackage{color}
\usepackage{amssymb}
\usepackage{amsmath}
\usepackage{dsfont}
\usepackage[colorlinks = false, citecolor=green, linkcolor=red]{hyperref} 

\newcommand{\ie}{i.e. \ } 
 
\newcommand{\sinc}{\mbox{sinc}}

\begin{document} 
 
\title{Analysis of quantum coherence in bismuth-doped silicon: a system of strongly coupled spin qubits} 
\author{M. H. Mohammady}
\affiliation{Department of Physics and Astronomy, 
University College London, Gower Street, London WC1E 6BT, United Kingdom} 
\author{G. W. Morley}
\affiliation{Department of Physics and Astronomy, 
University College London, Gower Street, London WC1E 6BT, United Kingdom}
\affiliation{London Centre for Nanotechnology
University College London, Gordon Street, London WC1H 0AH, United Kingdom}
\author{A. Nazir}
\affiliation{Department of Physics and Astronomy, 
University College London, Gower Street, London WC1E 6BT, United Kingdom}
\affiliation{Blackett Laboratory, Imperial College London, London SW7 2AZ, United Kingdom }
\author{T. S. Monteiro}
\affiliation{Department of Physics and Astronomy, 
University College London, Gower Street, London WC1E 6BT, United Kingdom}
\date{\today} 
\begin{abstract} 
  There is growing interest in bismuth-doped silicon  (Si:Bi) as an alternative to the well-studied proposals for silicon based quantum
 information processing (QIP) using phosphorus-doped silicon (Si:P). We focus here on the implications
of its anomalously strong hyperfine coupling. In particular, we analyse in detail the
regime where recent pulsed magnetic resonance experiments have demonstrated the potential for orders of magnitude
speedup in quantum gates by exploiting transitions that are electron paramagnetic resonance (EPR) forbidden at high fields. 
 We also present calculations using a phenomenological Markovian master equation which models the decoherence of the electron spin due
 to Gaussian temporal magnetic field perturbations. The model quantifies the advantages of certain
``optimal working points'' identified  as the $df/dB=0$ regions, where $f$ is the transition
frequency,  which come in the form of frequency minima
 and maxima. We show  that  at such regions, dephasing due to the interaction of the electron spin with a
 fluctuating magnetic field  in the $z$ direction (usually adiabatic) is completely removed.        
 
\end{abstract} 
\maketitle 
\section{Introduction}  
Beginning with the seminal proposal by Kane \cite{Kane}, there
 has been intense interest for over a decade in the use of Si:P 
\cite{SiP}  as qubits for quantum information processing. This donor-impurity spin-system
 continues to demonstrate an ever-increasing list of advantages for manipulation and storage of quantum information 
with currently available electron paramagnetic resonance (EPR) and nuclear magnetic resonance (NMR) technology. The Si:P system has four levels due to the electron spin $S=1/2$ coupled to a $^{31}$P nuclear spin $I=1/2$.

 The key advantages are the comparatively long decoherence times,  which have been measured to
be of order milliseconds for the electron spin of natural Si:P. They are of order seconds
for the nuclear spin, so the nuclear spin has been identified \cite{Kane} as a resource for storing the quantum information.
 For all but the weakest magnetic fields (i.e. $B_0 \gtrsim 200$G) \cite{Itoh}, the 
 electron and nuclear spins are uncoupled so may be addressed and manipulated independently
by a combination of microwave (mw) and radio-frequency (rf) pulses respectively. The two possible electron-spin transitions correspond to
EPR spectral lines, while the nuclear spin transitions are NMR lines.
Nuclear spin-flips are much slower: a $\pi$-pulse in the NMR case is   orders of magnitude longer than for the
EPR-allowed transitions.   

However, over the last year or so, there has also been increasing interest in another shallow donor impurity in silicon, the bismuth atom
\cite{LCN,Thewalt,Morton,Mohammady,Belli}. The
Si:Bi system is unique in several respects: it is the deepest donor, with a binding energy of about 71 meV, it has a very large
nuclear spin of $I=9/2$, it has an exceptionally large hyperfine coupling strength $A/2\pi=1.4754$ GHz.   $^{209}$Bi is the only naturally-occurring isotope. Recent measurements of the decoherence times in natural silicon
have revealed $T_2$ times of order $30\%$ larger than for Si:P, an effect attributed to the smaller Bohr radius of Si:Bi \cite{Morton}.
The dominant decoherence process is the spin diffusion \cite{DeSarma,YangLiu}, associated with the $I=1/2$,  $^{29}$Si nuclei 
occupying just under $5 \%$ of sites in
natural silicon; the dominant $^{28}$Si isotope has no nuclear spin and thus does not contribute to the dipole-coupled flip-flop process
which drives the spin diffusion.   A recent study of P donors in $^{28}$Si  purified to such a high degree  (less than 50 ppm of $^{29}$Si)
that spin diffusion may be neglected, revealed $T_2$ times potentially up to $10$ s \cite{Tyrishkin}. Although studies of isotopically
enriched Si:Bi  have  yet to be undertaken, since both species share the same $^{29}$Si decoherence mechanism,
$T_2$ times of the same order are to be expected.
 The coupling with $^{29}$Si was investigated in \cite{Belli}.
The very large nuclear spin $I=9/2$ and associated large Hilbert space may provide a means of storing more information \cite{LCN}.
Efficient hyperpolarization of the system (to about $90\%$) was demonstrated experimentally in \cite{Thewalt}.
  
The present study investigates  the implications of the very large  hyperfine coupling, $A/2\pi=1.4754$ GHz of Si:Bi, as well as its large nuclear spin.
Mixing of the Zeeman sublevels $|m_S,m_I\rangle$, achieved in the regime where the hyperfine coupling competes with
the external field, which we call the ``intermediate-field regime'', is not unexpected and has even been investigated
for Si:P for weak magnetic fields  \cite{Itoh}. However, for Si:Bi this regime is attained for magnetic fields  $B \simeq 0.1 - 0.6$ T which are
 moderate, but within the normal EPR range.  In a previous paper \cite{Mohammady}, we identified interesting consequences
 in this range of magnetic fields. Because the Rabi oscillation speed of a spin is dependent on its coupling strength to an external
 oscillating magnetic field, and the ratio of nuclear to electron coupling strengths is $2.488 \times 10^{-4}$, EPR pulses are
  orders of magnitude faster than NMR ones. We identified 
 a set of 4-states of Si:Bi,  which are, at high fields, entirely analogous to the 4-level subspace of 
Si:P. At the stated magnetic field range, all four possible transitions required for two-qubit universal quantum computation may be driven by fast EPR pulses (on a
nanosecond timescale), while in  Si:P two of the transitions require slow NMR pulses. 
 A recent experimental study using an S-band (4 GHz) pulsed EPR spectrometer \cite{Zurich} demonstrated the possibility of this strategy in Si:Bi.\\

 Furthermore, we identified a set of special magnetic field values: we show that here,  the effect of the external
field is wholly or partly cancelled by a component of the hyperfine interaction.
 We refer to this set of field values as ``cancellation resonances''.
 We show that the
cancellation resonance points are closely associated with with minima and maxima of the EPR spectral frequencies, (i.e. where $df/dB=0$).
We discuss an interesting  analogy with the electron
spin echo envelope modulation (ESEEM) phenomenon of ``exact cancellation'' \cite{Schweiger}, where like for the cancellation resonances, 
 the system Hamiltonian takes a simpler form. In exact cancellation, this leads to insensitivity to certain types
of ensemble averaging. In \cite{Mohammady}, it was found that an analogous insensitivity to ensemble averaging
over spin-exchange perturbations was seen at these points.

 Here, we investigate decoherence near the $df/dB=0$ points. In particular, we consider effects of
 Gaussian temporal magnetic-field fluctuations along the $x$ and $z$
direction on decoherence. We label these $X$ noise and $Z$ noise respectively.
 These may be relevant to the behavior of isotopically enriched Si:Bi. We show that for $Z$ noise, which is usually adiabatic,
 the $df/dB=0$ points offer decoherence-free zones. In analogy with work done on superconducting qubits \cite{Ithier},
 we call these ``optimal working points''. The system does not show such advantages for X noise, however, which leads to
 temperature-independent depolarising noise.

In section \ref{section2}  we follow our previous study \cite{Mohammady} by presenting
a full discussion of the spectral line
positions and transition strengths for coupled  nuclear-electronic spin systems for $S=1/2$ and
arbitrary $I$. We show that systems with large $A$ and $I$ display a rich structure of new EPR
transitions, many of which are forbidden at high fields (even as NMR transitions)
 and present a set of selection rules to classify 
four distinct types of transitions. We discuss the cancellation resonance points and explain their
relation to the maxima and minima of the transition frequencies.
In Section \ref{section3} we introduce the system as a pair of coupled qubits and compare with Si:P. 
We propose here a scheme of universal two-qubit quantum computation in the intermediate regime,
exploiting transitions forbidden at high field to obtain an orders of magnitude speed-up
relative to conventional Si:P qubits, which must combine fast EPR manipulation with much
slower NMR.
 In section \ref{section4} we introduce a model of decoherence caused by a
 temporal fluctuation of the external magnetic field, and study the effect of the cancellation resonances 
on the decoherence rates this model predicts. We conclude in Section \ref{section5}.

\section{Theory of coupled nuclear-electronic spin spectra}\label{section2}
\subsection{The Hamiltonian}
 \begin{figure}[tb] 
\includegraphics[width=3in]{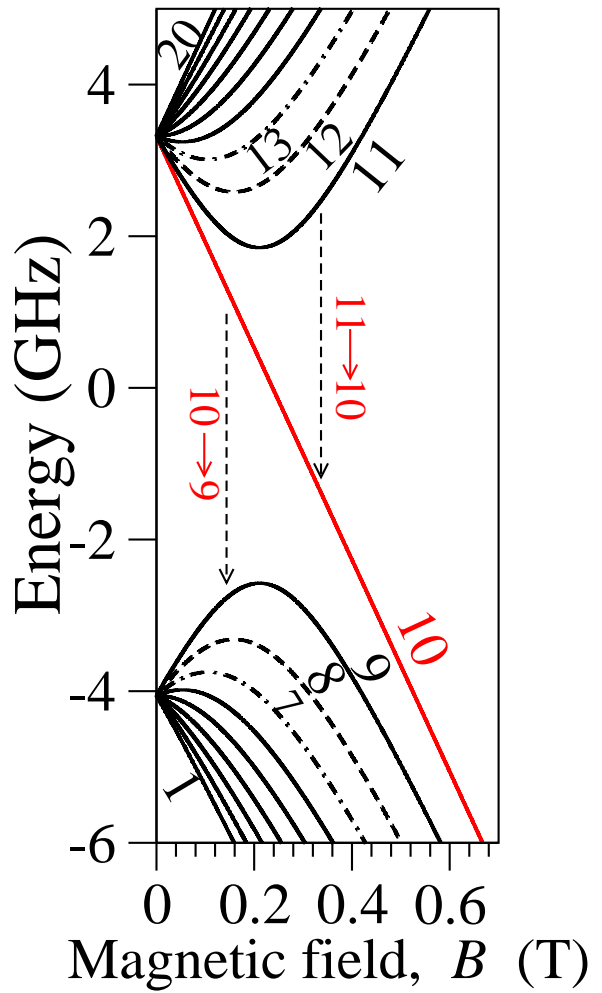} 
\caption{(Colour online) The 20 spin energy levels of Si:Bi may be labelled in  order of increasing energy $|1\rangle,|2\rangle ...|20\rangle$.
States $|10\rangle$ and $|20\rangle$ are not mixed. State $|10\rangle$ is of especial
significance since it, rather than the ground state, is a favourable state to 
initialise the system in. Experimental hyperpolarization studies \cite{Thewalt}
 concentrate the system in this state. Thus, in our coupled two-qubit scheme, state
$|10\rangle$ corresponds to our $|0_e0_n\rangle$ state; in the same scheme,
 states $|9\rangle \equiv |0_e1_n\rangle$ and  $|11\rangle \equiv |1_e0_n\rangle$
are related to $|10\rangle$ by a single qubit flip  while for $|12\rangle \equiv |1_e1_n\rangle $ both qubits are flipped.}
\label{Fig1} 
\end{figure} 

Nuclear-electronic spin systems such as Si:P and Si:Bi are described by the Hamiltonian:

\begin{equation} 
{\hat H_0}= \omega_0 {\hat S}_z - \omega_0 \delta {\hat I}_z  +  A {\bf{\hat S} . \bf{\hat I}} 
\label{HAM} 
\end{equation}    

\noindent where $\omega_0$ represents the electron Zeeman frequency given by $B g\beta$. Here, $B$ is the strength of the external magnetic field along the $z$ direction, $g$ is the electron g-factor, and $\beta$ is the Bohr magneton.   
$\delta=\omega_I/\omega_0 =2.488\times 10^{-4}$ represents 
the ratio of the nuclear to electronic Zeeman frequencies. $A$ is the isotropic hyperfine interaction strength. The operators $\bf{\hat S}$ and $\bf{\hat I}$ act on the  electronic and nuclear spins, respectively.

 For the systems considered,  the electron spin is always $S=1/2$. As a result the dimension of the Hilbert space is determined by the particular nuclear spin: for a given nuclear spin $I$, there 
are $2(2I + 1)$ eigenstates which can be superpositions of spin basis states $|m_S,m_I\rangle$.
However, since $[{\hat H_0},{\hat S}_z+{\hat I}_z]=0$, the Hamiltonian in Eq.\eqref{HAM} decouples to a direct sum of one and two-dimensional sub-Hamiltonians $ H_m^{1d}$ and $H_m^{2d}$ with constant $m=m_S+m_I$. The former act on the basis states $|m_S=\pm \frac{1}{2},m_I=\pm (I+\frac{1}{2})\rangle$, while the latter  act on the basis states $|m_S = \pm\frac{1}{2}, m_I=m \mp \frac{1}{2}\rangle$  such that $|m| < I+\frac{1}{2}$. The two-dimensional sub-Hamiltonians can be expanded in  the Pauli basis. In particular, the external field part of the  sub-Hamiltonian operator is given by:

\begin{equation} 
 \omega_0 {\hat S}_z- \omega_0\delta  {\hat I}_z  = \frac{\omega_0}{2}
\left[ (1 + \delta)\sigma_z -  2m \delta \mathds{1}\right].\label{externalfieldpart}
\end{equation}

\noindent The $z$ component of the hyperfine coupling,
\begin{equation} 
A {\hat S}_z \otimes {\hat I}_z =\frac{A}{2}\left( m  \sigma_z - \mathds{1}/2\right) \label{Isingpart}
\end{equation}
is seen to have an isotropic component as well as a non-isotropic component
dependent on ${ \sigma}_z$, while the $x$ and $y$ components are given by:
\begin{equation} 
A ({\hat S}_x \otimes {\hat I}_x +{\hat S}_y\otimes  {\hat I}_y)  = \frac{A}{2} \left[I(I+1)+\frac{1}{4}-m^2\right]^{1/2} { \sigma}_x .
\end{equation}

 \noindent Summing the above terms gives each  ${H}_m^{2d}$, whereas only Eqs. \eqref{externalfieldpart} and \eqref{Isingpart} contribute to  ${H}_m^{1d}$:

\begin{eqnarray} 
H_m^{2d} &=& \frac{A}{2}\left(\Delta_m{ \sigma}_z  + \Omega_m{ \sigma}_x   -\epsilon_m\mathds{1}\right) \nonumber \\H_{m=\pm(I+\frac{1}{2})}^{1d}&=& \frac{A}{2}(\pm\Delta_m-\epsilon_m) \nonumber \\
 \Delta_m &=& m+{\tilde \omega_0}(1+ \delta)\nonumber \\
\Omega_m &=& \left[I(I+1)+\frac{1}{4}-m^2\right]^{1/2}\nonumber \\
\epsilon_m &=& \frac{1}{2}(1+4{\tilde \omega_0}m \delta).
\label{Hm}
\end{eqnarray}
      
\noindent ${\tilde \omega_0}= \frac{\omega_0}{A}$ is the rescaled Zeeman frequency.
   We define a parameter 
$R_m^2 = \Delta_m^2 +\Omega_m^2$ where
$R_m$ represents the vector sum magnitude of  spin $x$ and $z$ components in the Hamiltonian.
We denote $\theta_m$ as the inclination of $R_m$ to the $z$-axis, such that
  $\cos \theta_m=\Delta_m/R_m$ and
 $\sin \theta_m=\Omega_m/R_m$. Then,
 $H_m^{2d}$  can also be written as   
\begin{eqnarray} 
H_m^{2d}= \frac{A}{2}\left(R_m\cos \theta_m {\sigma}_z +R_{m}\sin \theta_m{ \sigma}_x-\epsilon_m\mathds{1}\right.)\label{Hm1} 
\end{eqnarray} 
  
\noindent The range of values that $\theta_m$ can take are given by

\begin{eqnarray}
\theta_m\in\begin{cases}[0,\arctan\left(\frac{\Omega_m}{|m|}\right)] & \text{when} \ m > 0, \\ [0,\frac{\pi}{2}] & \text{when} \ m = 0,\\
[0,\frac{\pi}{2}+\arctan\left(\frac{\Omega_m}{|m|}\right)] & \text{when} \ m < 0, \\
\end{cases}
\end{eqnarray}

\noindent where the minimal value occurs as $B\to \infty$ and the maximal value occurs at $B=0$. Note that $\theta_m < \pi \ \forall \ B$.     
 
  Straightforward diagonalisation of  $H_m^{2d}$ then 
gives the eigenstates at arbitrary magnetic fields: 

\begin{equation} 
 |\pm, m\rangle = a_m^{\pm}|\pm\frac{1}{2}, m\mp\frac{1}{2} \rangle +  b_m^\pm  |\mp\frac{1}{2}, m\pm\frac{1}{2}\rangle,
\label{basis} 
\end{equation} 
\noindent where 
\begin{equation}
a_m^\pm = \cos\left(\frac{\theta_m}{2}\right) \ \ , \ \ b_m^\pm=\pm\sin\left(\frac{\theta_m}{2}\right)\label{coeffs}
\end{equation}

\noindent and with the corresponding eigenenergies:
\begin{equation} 
E^{\pm}_m = \frac{A}{2}\left[-\frac{1}{2}(1+4{\tilde \omega_0}m \delta) \pm R_m \right] .
\label{spectra} 
\end{equation} 
 
\noindent The high-field regime corresponds to $Bg\beta \gg A$. 
In this regime $\theta_m\to0$, hence $a_m^{\pm}\to 1$ and $b_m^{\pm}\to 0$;
the eigenstates in Eq.\eqref{basis} tend to the unmixed  $|m_S,m_I\rangle$ basis states. 
 The intermediate-field regime corresponds to $Bg\beta \sim A$ and strong mixing $|a_m^{\pm}|\sim |b_m^{\pm}|$. 
$H_m^{1d}$ has $\theta_m = 0 \ \forall \ m$, and hence gives the uncoupled eigenstates $|\pm\frac{1}{2},\pm(I+\frac{1}{2})\rangle$ at all magnetic fields.
 These have the simplified eigenenergies: 
\begin{equation}
E_{m=\pm(I +1/2)} =\pm\frac{\omega_0}{2}(1 - 2\delta I)+\frac{AI}{2}.\label{spectra1}
\end{equation}
It is important to stress that the $\sigma_z$ and $\sigma_x$ above are quite unrelated to the 
 $\hat S_z$ and $\hat S_x$  electronic spin operators. They are simply a method of representing the  two-dimensional sub-Hamiltonians.

In Fig.\ref{Fig1} the  exact expressions in  Eqs.\eqref{spectra} and \eqref{spectra1} were used to reproduce the spin spectra 
investigated for Si:Bi in for example, \cite{LCN} and \cite{Thewalt}.  These equations can be used to describe any arbitrary coupled nuclear-electronic spin system 
obeying the Hamiltonian Eq.\eqref{HAM}, such as other donor systems in Si including P and As. 
However, throughout this paper, we only present numerical solutions for Si:Bi. 
 As discussed here, its anomalously high value of $A$ and $I$ 
endows it with unique possibilities for spin based quantum computing. 

\subsection{Selection rules and transition strengths}
 The strength of EPR transitions between two spin eigenstates
may be characterised by a transition matrix element of typical form  $ |\langle\phi_i|\hat S_x|\phi_f\rangle|$,
where  the $|\phi_{i,f}\rangle$ are a pair of initial and final eigenstates involved in the transition.
At high fields  $ |\phi_{i}\rangle \equiv |m_S, m_I\rangle$ and the textbook selection rules $\Delta m_S =\pm 1, \Delta m_I=0$
determine which transitions are EPR-allowed and have non-zero transition intensity.
 In turn, NMR transitions have transition matrix element
 $ \delta |\langle\phi_i|\hat I_x|\phi_f\rangle|$ corresponding, at high fields,  to the selection rule
 $\Delta m_I =\pm 1, \Delta m_S=0$ for NMR-allowed transitions.
 The $\delta$ denotes the much weaker coupling between the nuclear magnetic dipole
and the external driving field, relative to the electronic spin-transitions typically observed in EPR spectroscopy.
Since $\delta \sim 10^{-4}$, this means that for typical, nanosecond-duration EPR driving pulses, one may
safely neglect the contribution of the much smaller $\hat I_x$ matrix element, when calculating spin qubit rotations.\\

However, in the intermediate field regimes, where
 $Bg\beta \sim A$, the eigenstates are strongly mixed. 
Then, transitions with non-zero $ |\langle\phi_i|\hat S_x|\phi_f\rangle|$ cannot be identified
by the familiar NMR or EPR selection rules.
Nevertheless, using the eigenstates in Eq.\ref{basis} we are able to
identify four types of transitions that can be observed at intermediate fields:
 $|\pm,m\rangle \leftrightarrow |\pm,m-1\rangle$ and $|\pm,m\rangle \leftrightarrow |\mp,m-1\rangle$.
 For a fixed $m$ the first two have transition frequencies, $\omega$,  that differ only by $2\delta \omega_0$, and
 similarly for the latter two.

 Transitions $|+,m\rangle \leftrightarrow |-,m-1\rangle$ are EPR-allowed for all magnetic fields,
 and their line intensities are proportional to
 \begin{equation} 
I^{+ \leftrightarrow -}_{m \leftrightarrow m-1} \propto |a^+_m|^2 |a^-_{m-1}|^2 = \cos^2\left(\frac{\theta_m}{2}\right)\cos^2\left(\frac{\theta_{m-1}}{2}\right). 
\label{Idip} 
\end{equation}

\noindent In the intermediate-field regime  $|+,m\rangle \leftrightarrow |+,m-1\rangle$ transitions (of intensity $I^+_{m\leftrightarrow m-1}$)
and  $|-,m\rangle \leftrightarrow |-,m-1\rangle$ transitions (of intensity $I^-_{m\leftrightarrow m-1}$), 
which are EPR-forbidden but NMR-allowed, at high field, now become EPR-allowed   
 with relative intensities:  
\begin{equation} 
I^{+}_{m \leftrightarrow m-1}\propto |a^+_m|^2 |b^+_{m-1}|^2= \cos^2\left(\frac{\theta_m}{2}\right)\sin^2\left(\frac{\theta_{m-1}}{2}\right) 
\label{Iforbp} 
\end{equation} 
\noindent and
\begin{equation} 
I^{-}_{m \leftrightarrow m-1 } \propto |a^-_{m-1}|^2 |b^-_{m}|^2=  \cos^2\left(\frac{\theta_{m-1}}{2}\right)\sin^2\left(\frac{\theta_{m}}{2}\right). 
\label{Iforbm} 
\end{equation}

\noindent One can see from Eq.\eqref{basis} that as $\omega_0 \to \infty$, the EPR intensities for these transitions goes as
  $I_{\pm, m\leftrightarrow m-1} \sim \frac{1}{\omega_0^2}   \to 0$ since  $|b^-_{m}|^2 \propto \frac{1}{\omega_0^2}$
at high fields. 

 However, the last transition type, namely $|-,m\rangle \leftrightarrow |+,m-1\rangle$, is most interesting
in that it is completely forbidden at high fields (it corresponds to neither an EPR-allowed nor
an NMR-allowed transition at high $B$) but nevertheless can correspond to significant transition strengths 
 at the intermediate-field regime. These are given by:

\begin{equation}
I^{- \leftrightarrow +}_{m \leftrightarrow m-1 } \propto |b^-_{m}|^2 |b^+_{m-1}|^2=  \sin^2\left(\frac{\theta_{m}}{2}\right)\sin^2\left(\frac{\theta_{m-1}}{2}\right). \label{Iforbidden}
\end{equation}

\noindent Clearly, such transitions never occur when the uncoupled eigenstates $|\pm,\pm(I + \frac{1}{2})\rangle$ are involved
as these eigenstates never exhibit mixing and always obey standard EPR or NMR selection rules.    
 
 \begin{figure}[!htb] 
\includegraphics[width=3.in]{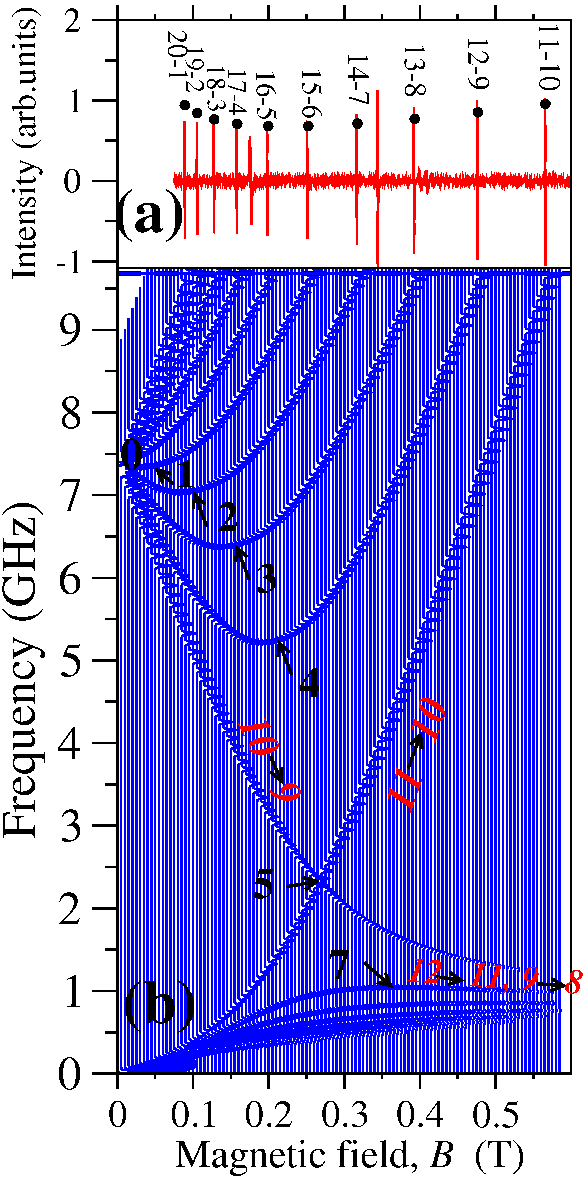} 
\caption{(Colour online) {\bf(a)} Comparison between theory [see Eqs.\eqref{spectra}, \eqref{spectra1} and \eqref{Idip}]
(black dots) and experimental CW EPR signal (red online) 
 at 9.7 GHz. Resonances without black dots above them 
are not due to Si:Bi; the large sharp resonance at $0.35$ T is due to
silicon dangling bonds, while the remainder are due to defects in the
sapphire ring used as a dielectric microwave resonator. 
The variation in relative intensities is mainly due to the mixing of states as 
in Eq.(\ref{basis}). The variability is
not too high but the  calculated intensities are consistent with
experiment and there is excellent agreement for the line positions. 
 {\bf (b)} Calculated EPR spectra
(convolved with a 0.42 mT measured linewidth); they are seen to line-up with the
experimental spectra at $f=\omega/2\pi=9.7$ GHz).
The  type I cancellation resonances are indicated by  integers $-m=0,1,2,3,4,5$. The first four of these are associated with $df/dB=0$ points.
 The type II cancellation resonance at $\tilde\omega_0\simeq7$ also coincides with a $df/dB=0$ point, and  corresponds to
that shown in the $\lesssim 2$ GHz electron nuclear double resonance (ENDOR) spectra of \cite{Morton}.} 
\label{Fig2} 
\end{figure}

\subsection{Cancellation resonances}

As shown above,  the constant $m$ states of Eq.\ref{basis}  are eigenstates of the
Hamiltonian 
$H_m^{2d} = \frac{A}{2}\left(\Delta_m{ \sigma}_z  + \Omega_m{ \sigma}_x\right)$ -- given by Eq.\ref{Hm} excluding a trivial shift --
where $\Delta_m \simeq  m+{\tilde \omega_0}$. For the Si:Bi spectra
of  Fig.\ref{Fig1},  this encompasses nine pairs
 of states (i.e. all except the uncoupled states $|10\rangle$ and $|20\rangle$, which are governed by the $H_m^{1d}$). We use the term ``cancellation resonance'' as a blanket term for magnetic field regimes that  simplify the system Hamiltonian. There are two types of cancellation resonance:

\begin{description}
\item[Type I] :
 $\Delta_m$=0, taking place when $\tilde \omega_0 \simeq  -m$
\item[Type II]:  $\Delta_m=\Omega_m$, taking place  when $\tilde\omega_0\simeq-m+\Omega_m$ 
\end{description}

\noindent In the Si:Bi system,  with $I=9/2$, the  type I cancellation resonance  corresponds to
 $m=0,-1,-2,-3,-4,-5$ and a set of equally spaced magnetic field values $B=0,0.05,...,0.21,0.26$ T. For $-4 \leq m\leq 0$, the  term in $H_m^{2d}$ dependent on $\sigma_z $  vanishes entirely at the cancellation resonance.
These are associated with Landau-Zener crossings.
The point at which $m+{\tilde \omega_0} \simeq 0$ for $m=-(I+\frac{1}{2})$ also has special interest
(see below) although it is not a Landau-Zener crossing. Here  $\Omega_m=0$ too.
For Si:Bi it corresponds to $m=-5$ and $B=0.26$T.

The  type II cancellation resonance  is particularly interesting for the $m=-3,-4$ subspaces, where at $\tilde \omega_0\simeq7$ we have   ${H}_{m=-3,-4}^{2d} \propto (\sigma_x+\sigma_z)$
(ignoring the trivial term proportional to the identity). 
Although the term cancellation resonance is simply a
convenient label, the type I variant is somewhat reminiscent of the ESEEM phenomenon of
exact cancellation; here too , the $\sigma_z$ component of the Hamiltonian  
vanishes, leading to insensitivity to ensemble averaging. Thus we briefly discuss
the parallels below.

\subsection{Analogy with ``exact cancellation''}
Exact cancellation is a widely used ``trick'' in ESEEM spectroscopy. A
coupled nuclear-electronic system with anisotropic hyperfine coupling, 
 which is weak compared with
electron spin frequencies (on the MHz scale rather than GHz scale), has
a rotating frame Hamiltonian \cite{Schweiger}:

\begin{equation} 
{\hat H}_0= \Omega_s {\hat S}_z + \omega_I  {\hat I}_z  + A_1 {\hat S}_z \otimes {\hat I}_z +
                           A_2 {\hat S}_z \otimes {\hat I}_x .
\end{equation}

\noindent Here $\Omega_s=\omega_0-\omega$ is the detuning from the external driving field
and $\omega_I= \delta \omega_0$ is the nuclear Zeeman frequency.
$A_1$ and $A_2$ are  secular and pseudo-secular 
hyperfine couplings as given in standard texts \cite{Schweiger}. At resonance,
$\Omega_s=0$. As the hyperfine terms are weak, terms like ${\hat S}_x\otimes  {\hat I}_x+{\hat S}_y \otimes {\hat I}_y $
are averaged out by the rapidly oscillating (microwave) driving. The remaining Hamiltonian $ \omega_I  {\hat I}_z  + A_1 {\hat S}_z \otimes {\hat I}_z + A_2 {\hat S}_z \otimes {\hat I}_x$
conserves $m_S$. For a spin $S=\frac{1}{2}, I=\frac{1}{2}$ system like Si:P, the Hamiltonian decouples into
two separate $2\times 2$ Hamiltonians ${\hat H}_{m_S=\pm\frac{1}{2}}$.
 In the $m_S=+\frac{1}{2}$ subspace,

\begin{equation} 
{\hat H}_{m_S=+\frac{1}{2}}=  \frac{1}{2}\left(\omega_I + \frac{A_1}{2}\right){ \sigma}_z+ \frac{A_2}{2}{ \sigma}_x
\label{plus}
\\ \end{equation}

\noindent where the Pauli matrices are defined relative to the basis
 $|m_S\rangle\otimes |m_I\rangle= |+\frac{1}{2}\rangle\otimes |\pm \frac{1}{2}\rangle$ 
while in  the $m_S=-\frac{1}{2}$ subspace,
\begin{equation} 
{\hat H}_{m_S=-\frac{1}{2}}=  \frac{1}{2}\left(\omega_I - \frac{A_1}{2}\right){ \sigma}_z- \frac{A_2}{2}{ \sigma}_x
\label{minus}
\end{equation}
where the Pauli matrices are defined relative to the basis
 $|m_S\rangle\otimes |m_I\rangle= |-\frac{1}{2}\rangle\otimes|\pm \frac{1}{2}\rangle$.
It is easy to see from Eq.(\ref{minus}) that if 
$ \omega_I = A_1/2$,
\noindent only the $A_2\sigma_x/2$ term remains. This is the ``exact cancellation''
condition. While reminiscent of hyperfine cancellations resonances,
 there are key differences.
In particular, since the type I cancellation resonances of Eq.(\ref{Hm})
 affect both nuclear and electron spins, 
at $m\simeq-{\tilde \omega_0}$, the eigenstates assume ``Bell-like'' form:

\begin{equation} 
|\Psi^\pm \rangle=  \frac{1}{\sqrt{2}} \left( |-\frac{1}{2}\rangle_e\otimes|m+\frac{1}{2}\rangle_n \pm
 |+\frac{1}{2}\rangle_e \otimes|m-\frac{1}{2}\rangle_n \right)
\end{equation}

\noindent where the $e,n$ subscripts have been added for clarity, to indicate the electronic
and nuclear states respectively. In contrast,
 for exact cancellation, they give superpositions of nuclear spin states only:
\begin{equation}
|\Psi \rangle= \frac{1}{\sqrt{2}} |-\frac{1}{2}\rangle_e\otimes \left(|+\frac{1}{2}\rangle_n \pm |-\frac{1}{2}\rangle_n\right),
\end{equation}
 
\noindent which still permits interesting manipulations of the nuclear spin states \cite{Mitrikas}.

Note that, while exact cancellation eliminates the full Ising term $A_1 {\hat S}_z \otimes {\hat I}_z$,
the EPR cancellation resonance eliminates only the non-isotropic part. 
 Furthermore, as discussed above, cancellation resonances also have a type II variant. The $\tilde\omega_0=7$  resonance does
not cancel the hyperfine coupling at all; it equalizes the Bloch vector of
the states in adjacent m-subspaces producing another effect.

 EPR cancellation resonances are in practice a much stronger effect than exact cancellation: 
decohering and perturbing effects of interest in quantum information
 predominantly affect the electronic spins, not the nuclear spins. Exact cancellation appears in 
 the rotating frame Hamiltonian (which contains only terms of order MHz). It will not
survive perturbations approaching the GHz energy scale.
 The cancellation resonances, on the other hand,
 arise in the full Hamiltonian, eliminate large electronic terms and can potentially
thus reduce the system's sensitivity to major sources of broadening and decoherence.

It is valuable to recall a major reason why the ``exact cancellation'' regime is so widely
exploited in spectroscopic studies. In systems with anisotropic coupling, the spectra
depend on the relative orientation of the coupling tensor and external field. Thus for
powder spectra, which necessarily average over many orientations,
 very broad spectral features result. At exact cancellation, the simplification
of the Hamiltonian is dramatically signalled by ultra-narrow spectral lines \cite{Schweiger}.
Similarly, in \cite{Mohammady}, insensitivity to perturbation by a spin ensemble, in the 
form of ultra-narrow spectral lines, was demonstrated in the cancellation resonance regime.
This motivates further investigation of the potential of the cancellation
resonance points for reducing decoherence. 

\subsection{The frequency minima and maxima}
 
In Fig.\ref{Fig2}, we show Si:Bi spectra in the intermediate field regime, using the expressions
for frequencies and transition strengths presented above. In Fig.\ref{Fig2}(a) we show a 
comparison with experimental spectra, showing good agreement with line intensities and 
positions.
A striking feature of Fig.\ref{Fig2}(b) are a set of spectral minima and maxima
of the transition frequency of several lines.
These are close, but not coincident with the type I cancellation resonance points (indicated
by arrows and labelled $-m=0,1,...$); for instance, while the $-m=4$ cancellation resonance
corresponds to an avoided crossing between states $|11\rangle$ and $|9\rangle$,
and the  $-m=3$ point corresponds to an avoided crossing between states $|12\rangle$ and $|8\rangle$,
 the nearby frequency minimum involves the transitions $|12\rangle \leftrightarrow |9\rangle$ and $|11\rangle \leftrightarrow |8\rangle$. In other 
words it involves two states from adjacent avoided crossings.
This rich EPR structure is entirely absent in (say) conventional Si:P spin systems (with $I=1/2$ and small $A$),
which do not have these multiple avoided crossings, at quite high magnetic fields.
It is nonetheless possible to fully analyse this structure for Si:Bi without resorting to numerics.

 We can show that  transitions of type $|\pm,m\rangle \leftrightarrow |\mp,m-1\rangle$  have a unique $B$ value for which  $df/dB=0$     when
 \begin{equation}
  \cos(\theta_m)\simeq-\cos(\theta_{m-1})
 \end{equation} 
  \noindent if $-I+\frac{3}{2} \le  m \le 0$.  Such a condition can only be satisfied if $\theta_m \sim\theta_{m-1} \sim\pi/2$, meaning that both states must be near a Landau-Zener type cancellation resonance. The value of $B$ which satisfies this is 

\begin{equation}
B \simeq -\frac{A}{g\beta}\frac{(m-1)\Omega_m   + m\Omega_{m-1}  }{\Omega_{m-1} + \Omega_m}.
\end{equation} 

\noindent  
Further study of these $df/dB=0$ points show that they are frequency minima, and they can be observed in Fig.\ref{Fig2}(b) near the cancellation resonance points marked ``0,1,2,3,4''.
An equivalent way of viewing the frequency minimum condition
$ \cos \theta_m \simeq  -\cos \theta_{m-1} $
\noindent is to write:

\begin{eqnarray}
\theta_m &\simeq& \frac{\pi}{2}-\phi \nonumber \\
\theta_{m-1} &\simeq& \frac{\pi}{2} + \phi
\end{eqnarray}
so the frequency minima occurs when both subspaces involved are an equal ``{\em angular distance}'' away
 from their cancellation resonance points. 
\\

Transitions  $|\pm,m\rangle \leftrightarrow |\pm,m-1\rangle$ also have a $df/dB=0$ point when

\begin{equation}
\cos \theta_m \simeq \cos \theta_{m-1}\label{max}
\end{equation}

\noindent if $-I+\frac{3}{2} \leq m \leq 0$.  These $df/dB=0$ points are frequency maxima and are given at fields: 

\begin{equation}
B \simeq \frac{A}{g\beta}\frac{(m-1)\Omega_m   - m\Omega_{m-1}  }{\Omega_{m-1} - \Omega_m}.
\end{equation}

\begin{figure}[!htb] 
\includegraphics[width= 3. in]{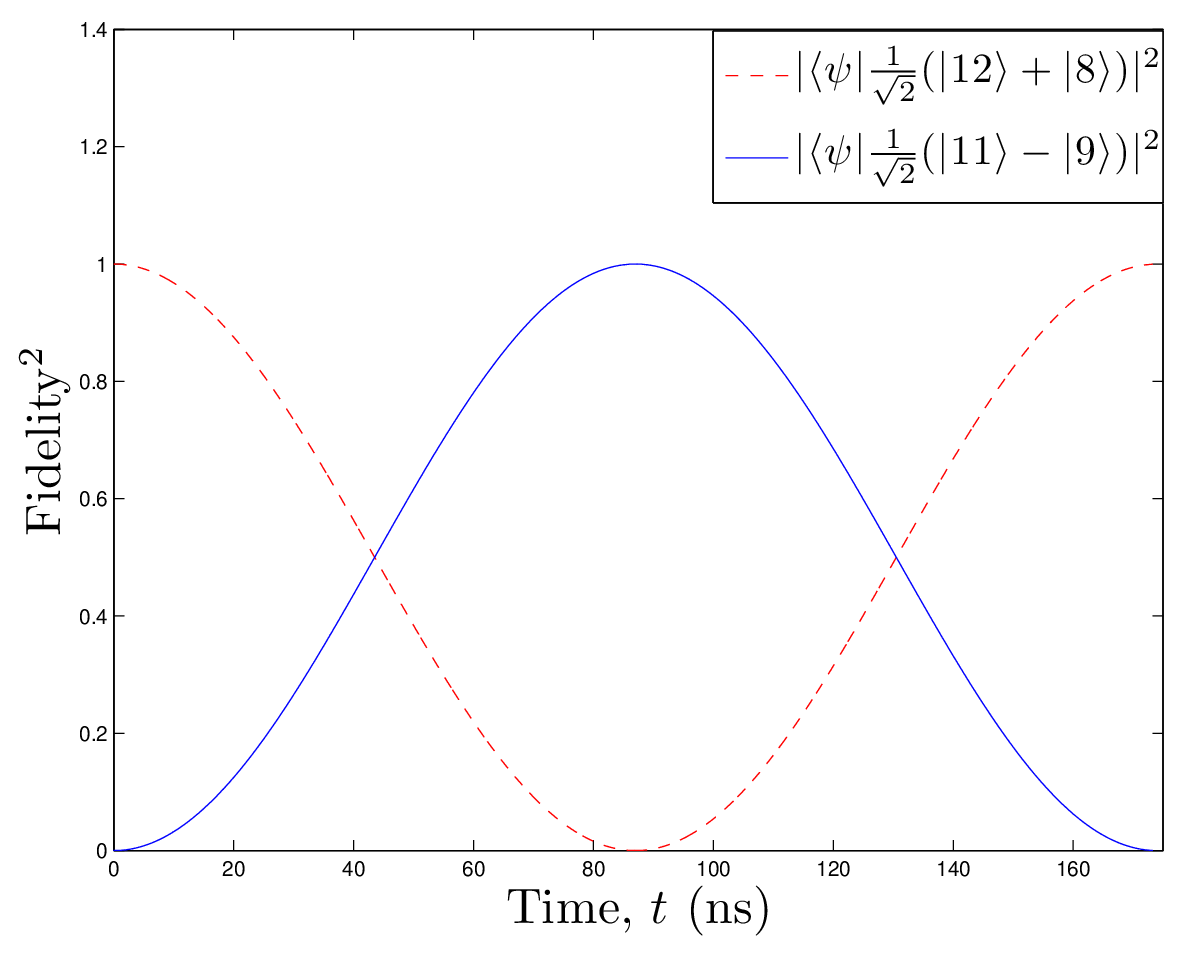} 
\caption{ (Colour online) Shows that near the ${\tilde \omega_0}=7$ frequency maximum, the transition rates $|12\rangle \leftrightarrow |11\rangle$ and $|8\rangle \leftrightarrow |9\rangle$ equalise, and we may transfer the coherences between the former to the latter, with a relative phase shift of $\pi$. We use $\omega_1/2\pi=200$ MHz.}   
\label{Fig3} 
\end{figure}

\noindent Because  $0\leq\theta_m < \pi$, the frequency maximum condition Eq.\eqref{max}  implies that $\theta_m \simeq\theta_{m-1}$. In the case of Si:Bi
only the maximum for the transitions $|\pm,-3\rangle\leftrightarrow|\pm,-4\rangle$ occurring at ${\tilde \omega_0}\simeq 7$ and $B\simeq 0.37$ T  can be observed by EPR [this is shown in the region of Fig.\ref{Fig2}(b) labeled ``7'']. The other maxima occur at fields $B > 0.5$ T for which the EPR line intensities become vanishingly small.  
The ${\tilde \omega_0}\simeq 7$ frequency maximum is especially interesting because at this value both the $m=-3,-4$ subspaces are at their type II cancellation resonance. Here,
 $\theta_{-3}\simeq \theta_{-4}\simeq\pi/4$, which implies that $H_{-3}^{2d} \propto H_{-4}^{2d} \propto (\sigma_x+\sigma_z)$. 
Such a symmetrisation of the Hamiltonian  offers possibilities for more complex manipulations. It has been suggested 
\cite{Morton,LCN} that the larger state-space of Si:Bi may be used to store
more information. Thus  we can show  that at $\tilde\omega_0 \simeq 7$, a single
 EPR ($\sim 80$ ns)  pulse can map any coherences
 between the $m=-4$ states  into the same coherences between the $m=-3$ states.
 The condition Eq.\eqref{max}  implies that the amplitudes $a^\pm_{-3}\simeq a^{\pm}_{-4}$ 
and $b^\pm_{-3}\simeq \pm b^{\pm}_{-4}$. This means that an  EPR pulse
  will effect the rotations $|12\rangle\leftrightarrow|11\rangle$  and   $|9\rangle \leftrightarrow |8\rangle$ at the same rate. 
 For instance, if the initial two-qubit state is
 $|\Psi\rangle= c_{11} |12\rangle +c_{9} |8\rangle$  a $\pi$-pulse
 will yield $|\Psi\rangle= c_{11} |11\rangle - c_{9} |9\rangle$, and 
so produces a mechanism for temporarily storing the two-qubit state (within a 
relative $\pi$ phase shift). This is illustrated in Fig.\ref{Fig3}.\\

 \begin{figure}[!htb] 
\includegraphics[width=3.0 in]{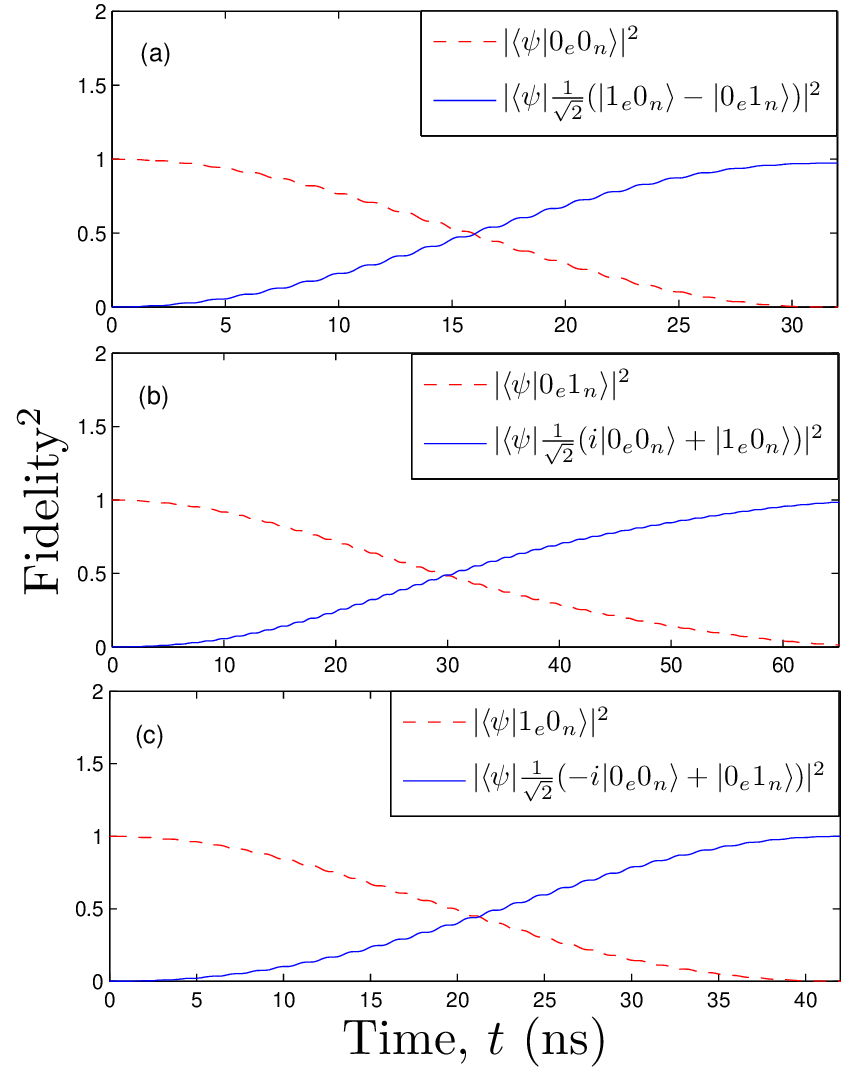} 
\caption{ Shows that at the ${\tilde \omega_0}=5$ resonance, second order,
two-photon transitions may be exploited since $f^{10\leftrightarrow9} \simeq  f^{11\leftrightarrow10}$. A linear oscillating microwave field of strength $\omega_1/2\pi = 200$ MHz is used.
{\bf (a)} shows that driving at resonance prepares $|10\rangle \to \frac{1}{\sqrt{2}}(|11\rangle -|9\rangle)$.
The process is very sensitive to detuning from resonance. {\bf (b)} and {\bf (c)} illustrate
how slight detuning of the microwave frequency may be used to prepare other superpositions such 
as $ |9\rangle \to \frac{1}{\sqrt 2} \left(i|10\rangle +|11\rangle\right)$ 
and $ |11\rangle \to \frac{1}{\sqrt 2} \left(-i|10\rangle +|9\rangle\right)$.}   
\label{Fig4} 
\end{figure}
  
  The $m=-5$ state of Si:Bi,  $|10\rangle$,
 is not  associated with a Landau-Zener crossing
 at any field as the Hamiltonian leaves it uncoupled to any other basis state.
 Nevertheless the fields for which
$ {\tilde \omega_0} = 5(1+\delta)$ (at $B \approx 0.26$ T for Si:Bi)
represent the most drastic case of  type I cancellation resonance: the $\Delta_{-5}$ term in $H_{-5}^{1d}$ vanishes, leaving only the   $\epsilon_{-5}$  term. Here
 $E_{-5}\simeq-A/4$, so its energy lies almost exactly half-way between 
the $|\pm,-4\rangle$ state energies: states $|9\rangle$ and $|11\rangle$ of Si:Bi have energies
$E_{\pm}\simeq E_{-5} \pm R_{-4}$. This gives the striking feature
at $  2.3$ GHz in Fig.\ref{Fig2}(b) where 
the $|10\rangle \leftrightarrow |9\rangle$ and $|11\rangle \leftrightarrow |10\rangle$ lines coincide and where
an EPR pulse would simultaneously generate coherences between state $|10\rangle$
 and {\em both} states $|11\rangle$ and $ |9\rangle$. In effect one may use two-photon,
second order processes to transfer population between states $|9\rangle$ and $|11\rangle$ 
(recall that  simultaneous spin flips are forbidden for isotropic hyperfine
coupling). Fig.\ref{Fig4} illustrates this.

\
\section{Si:Bi as a two-qubit system}\label{section3}
\subsection{Basis states}

The adiabatic eigenstates of the well-studied four-state $S=1/2$, $I=1/2$ Si:P system can be mapped
onto a two-qubit computational basis, as shown in Table \ref{sipqubitstates}.

\begin{table}[!htb]
\begin{tabular}{c|c|c}
Adiabatic state   & high-field state &  logical qubit  \\
\hline
 $|4\rangle $ & $|+\frac{1}{2},+\frac{1}{2}\rangle$   &$|1_e1_n\rangle$\\
 $|3\rangle $ & $|+\frac{1}{2},-\frac{1}{2}\rangle$   &$|1_e0_n\rangle$\\
 $|1\rangle$ & $|-\frac{1}{2},+\frac{1}{2}\rangle$   &$|0_e1_n\rangle$\\
 $|2\rangle$ & $|-\frac{1}{2},-\frac{1}{2}\rangle$   &$|0_e0_n\rangle$\\
 
\hline
\end{tabular}
\caption{Two-qubit computational basis states of Si:P}\label{sipqubitstates}
\end{table}

\noindent With a 20-dimensional state-space, the Si:Bi spectrum is considerably more
complex. However, we can identify a natural subset of 4 states
(states $|9\rangle,|10\rangle,|11\rangle$ and $|12\rangle$), which
represents an effective coupled two--qubit analogue, as shown in Table \ref{sibiqubitstates}.
As hyperpolarization initialises the spins in state 
$|10\rangle $ \cite{Thewalt} and this state has both the electron and nuclear spins
fully anti-aligned with the magnetic field, although it is not the ground state,
 it can be identified with the $|0_e0_n\rangle$ state. The other states 
-- just as in the Si:P case -- are related to it by adding a single quantum of spin to
one or both qubits.

\begin{table}[!htb]
\begin{tabular}{c|c|c}
Adiabatic state   & high-field state &  logical qubit  \\
\hline
 $|12\rangle $ & $|+\frac{1}{2},-3\frac{1}{2}\rangle$   &$|1_e1_n\rangle$\\
 $|11\rangle $ & $|+\frac{1}{2},-4\frac{1}{2}\rangle$   &$|1_e0_n\rangle$\\
 $|9\rangle$ & $|-\frac{1}{2},-3\frac{1}{2}\rangle$   &$|0_e1_n\rangle$\\
 $|10\rangle$ & $|-\frac{1}{2},-4\frac{1}{2}\rangle$   &$|0_e0_n\rangle$\\
 
\hline
\end{tabular}
\caption{Two-qubit computational subspace of Si:Bi}\label{sibiqubitstates}
\end{table}

For both systems, there are, in principle, four transitions that would account for all possible individual qubit operations, as   listed in Table \ref{table1}. We show below that for
Si:Bi, all  qubit  operations are EPR-allowed for $B \sim 0.1 - 0.6$ T.
For Si:P, this region permits EPR-manipulation of only the electronic qubit-flips (the first two); 
nuclear rotations require much slower, $\mu$s, NMR transitions. Measurement of the qubits in the computational basis has to be performed at high fields, where the adiabatic logical qubit coincides with the electron and nuclear spin states.
All simultaneous nuclear and electronic qubit flips are  forbidden for systems with isotropic
hyperfine coupling $A$ including both Si:P and Si:Bi.
We note that, in spin-systems with ``exact cancellation'' and anisotropic $A$,
the $A \hat I_x \otimes\hat S_z$ coupling does permit simultaneous nuclear-electronic qubit flips. These were
recently shown for the organic molecule malonic acid \cite{Mitrikas}; the disadvantage here is that
single nuclear qubit rotations (essential for quantum computation) are not EPR-allowed.

\subsection{Universal set of quantum gates}

\begin{table}[!tb]
\begin{tabular}{c|c|c}
Controlled operation   & Si:P transitions &  Si:Bi transitions  \\
\hline
 $\hat R_{\underline{v}}(\theta)_e\otimes|0\rangle\langle 0|_n $ & ${\bf \omega^{3\leftrightarrow2}}$   &${\bf \omega^{11\leftrightarrow10}}$\\
 $\hat R_{\underline{v}}(\theta)_e\otimes|1\rangle\langle 1|_n $ & ${\bf \omega^{4\leftrightarrow1}}$   &${\bf \omega^{12\leftrightarrow9}} $\\
 $|0\rangle\langle0|_e\otimes\hat R_{\underline{v}}(\theta)_n$ & $    {\omega^{2\leftrightarrow1}}$   &${\bf \omega^{10\leftrightarrow9}}$\\
 $|1\rangle\langle1|_e\otimes\hat R_{\underline{v}}(\theta)_n$ & $    {\omega^{4\leftrightarrow3}}$   &${\bf \omega^{12\leftrightarrow11}}$\\
 
\hline
\end{tabular}
\caption{conditional single-qubit rotations of angle $\theta$ about vector $\underline{v}$ in the Bloch sphere, denoted $\hat R_{\underline{v}}(\theta) $, and corresponding transition frequencies. 
Frequencies in boldface correspond to qubit operations
which are EPR-allowed at $B=0.1-0.6$ T; \ie they require only  fast (ns) EPR pulses. 
All four EPR  operations are possible for Si:Bi, whereas  for Si:P
nuclear qubit operations require  slow ($\mu$s) NMR pulses. This scheme allows for cheap, controlled  qubit operations, whereas single qubit operations would require twice the number of pulses.}
\label{table1}
\end{table}

It is known that for universal quantum computation it suffices to be able to perform arbitrary single qubit
 rotations and a two-qubit gate such as the CNOT \cite{DiVincenzo}. We now show how we may exploit the strong 
hyperfine interaction of the Si:Bi system to achieve this using only fast EPR pulses, eliminating the
need for the much slower (longer-duration) NMR pulses. 

Control of the electron spins is facilitated by the  
Hamiltonian ${\hat H}={\hat H_0} + V_{x/y}(t)$ where
 $V_{x/y}(t) =\omega_1\cos( \omega t) \hat{S}_{x/y} $ represents the external magnetic field oscillating  along the $x$ or $y$ axis. This may be written as:

\begin{eqnarray}
V_{x/y}(t) &=& \frac{\omega_1}{2}[\cos(\omega t)\hat S_{x/y} + \sin(\omega t)\hat S_{y/x}] + 
\nonumber \\ &&\frac{\omega_1}{2}[\cos(\omega t)\hat S_{x/y} - \sin(\omega t)\hat S_{y/x}].
\label{Vxyp}\end{eqnarray}

 \noindent We label the first component the right handed (RH), and the second term the left handed (LH) rotating fields.
 In the rotating frame between two eigenstates $|e\rangle$ and $|g\rangle$, which satisfy the selection rule $|\langle e|\hat S_{x/y}|g\rangle| =|\eta|   > 0$, the
 Liouville-von Neumann equation for the reduced two-level system is
\begin{equation}
 \frac{d{ \tilde\rho(t)}}{d t}= i\frac{\omega_1\eta}{4} [\tilde\rho(t),\hat\sigma_{x/y} ]\label{vonneumann}
\end{equation} 

\noindent if $\omega_1 \ll \omega^{e\leftrightarrow g}$, where $\omega^{e\leftrightarrow g}$ is the transition frequency between the two eigenstates.  For the transitions where the increase in energy corresponds to an 
increase(decrease) in total $z$-axis magnetisation, $m$, the resonance condition is satisfied by
 the RH(LH) component of the oscillating magnetic field . This feature, which  is explained in more detail
 in Appendix.\ref{unitarydynamics}, may be exploited for qubit manipulation involving certain transitions which are
near-degenerate, as  will be explained in the following section. 

The EPR pulses at our disposal allow us to perform controlled single qubit unitaries $R_{\underline{v}}(\theta)$ where $\underline{v}$ lies in the $x-y$ plane.
 Two orthogonal Paulis suffice to generate  arbitrary single-qubit unitaries \cite{NielsenChuang} using at most three pulses, 
  and we may construct the controlled $\hat\sigma_z$  and Hadamard gates  by these pulse sequences:

\begin{eqnarray}
\hat\sigma_z&=&e^{i \frac{3\pi}{2}}e^{-i\frac{\pi}{2}\hat\sigma_y}e^{-i\frac{\pi}{2}\hat\sigma_x},\nonumber \\ H:=\frac{1}{\sqrt{2}}(\hat \sigma_x + \hat \sigma_z) &=& e^{i \frac{3\pi}{2}}e^{-i\frac{3\pi}{4}\hat\sigma_y}e^{-i\frac{\pi}{2}\hat\sigma_x}.
\end{eqnarray}

\noindent The possible controlled operations are shown in Table \ref{table1}. Single-qubit gates would require us to repeat the set of controlled EPR pulses for both the controlling qubit basis states, and as such would require twice the time.

The  transition strengths given in Eqs.(\ref{Idip}), (\ref{Iforbp}), (\ref{Iforbm}) and (\ref{Iforbidden}) are given by  $|\eta|^2$. Eq.(\ref{vonneumann}) shows that the qubit rotation speed, given a fixed microwave field strength, is determined by the mixing factor $\eta$.  As $B\to\infty$, $\eta\to1$ for high-field EPR transitions, and $\eta\to0$ for high-field NMR transitions as well as the high-field dipole-forbidden transition. At magnetic fields where $A \sim  B g \beta$, however,
 mixing occurs and $\eta$ will become appreciable.

  At the $m=-4$ cancellation resonance, corresponding to field values $\tilde \omega_0\simeq4$ ($B \simeq 0.21$ T),
the values of  $|\eta|$ for both nuclear and electronic qubit operations equalise:
 this is simple to verify from Eqs.(\ref{Idip})-(\ref{Iforbm}) by setting $\theta_{-5}=0$ and $\theta_{-4}=\pi/2$.
 We show numerically in Fig.\ref{Fig5}(a)-(b) 
that this means a $\pi$ pulse on the nuclear qubit in effect becomes as short as on
the electronic one.

\subsection{Selective qubit gates for near-degenerate transitions}

An important advantage of using the Si:Bi at intermediate fields is the prospect of 
quantum computing using exclusively fast, nanosecon EPR pulses. Nevertheless, such short
pulses necessarily imply a larger frequency bandwidth.
While this is not, in general, a problem, it may present difficulties for
certain pairs of transitions that are quite close in frequency (tens of MHz rather than
GHz): the EPR pulse may drive unwanted spin flips.  One solution is to simply lengthen the duration of
the pulse; however one then loses much of the speed-up advantage as timescales comparable to
NMR are then required. We show here that it remains possible to perform selective
one-qubit gates with fast EPR pulses.

\begin{figure}[!htb] 
\includegraphics[width=3.6 in]{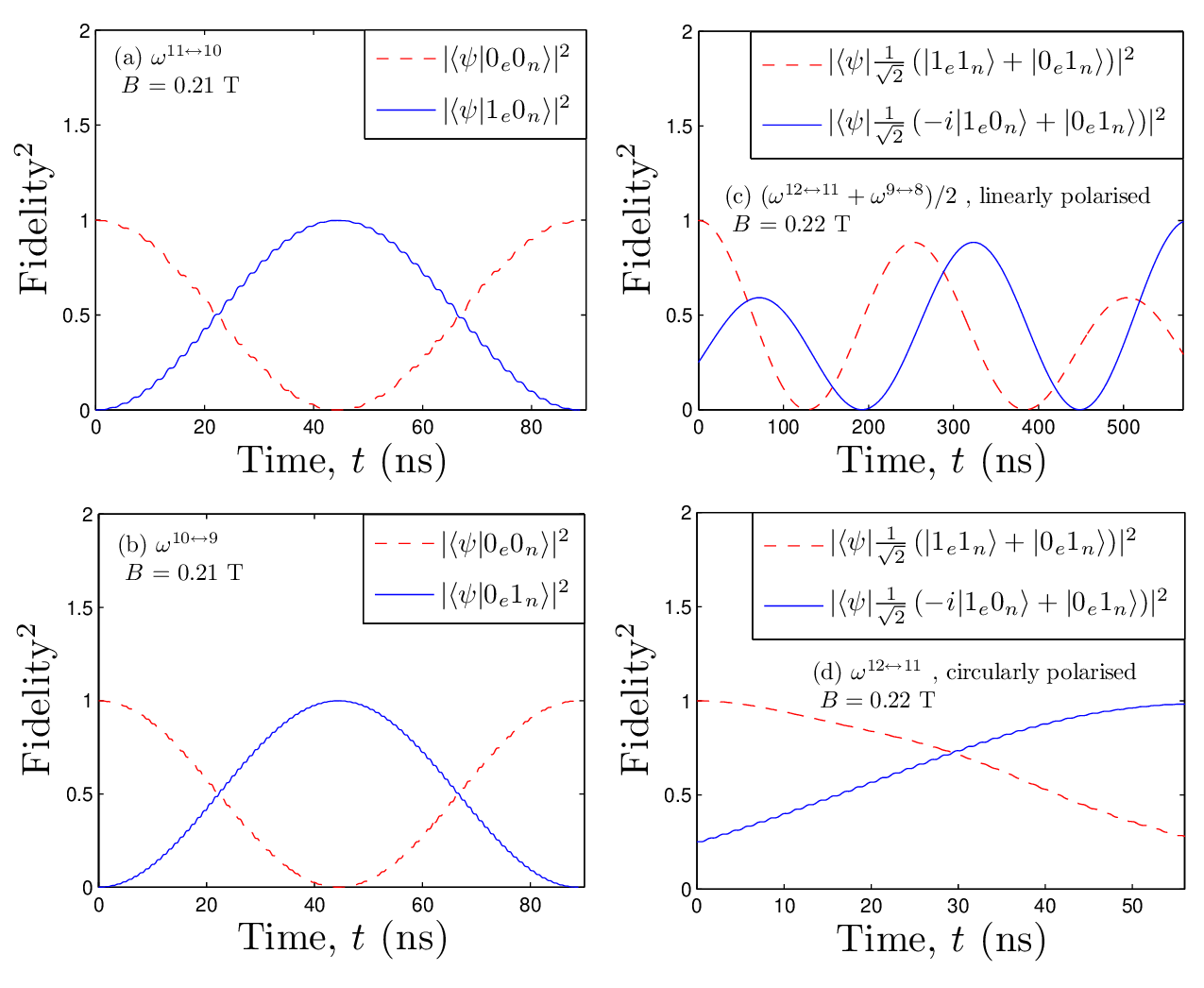} 
\caption{(Colour online){\bf (a)} and {\bf (b)} show Rabi oscillations utilising a linearly oscillating microwave field of strength $\omega_1/2\pi = 200$ MHz.  
 At $B =0.21$ T,  the time taken for  the electron qubit flip $|0_e0_n\rangle\to|1_e0_n\rangle$
is identical to that for the nuclear qubit flip $|0_e0_n\rangle\to|0_e1_n\rangle$. {\bf (c)} and {\bf (d)} show selective nuclear qubit flips, and have microwave fields of strength $\omega_1/2\pi= 100$ MHz. This is because the energy difference  between the eigenstates is smaller than in the previous two cases and $\omega_1$ must remain perturbative. {\bf (c)} Utilises a linearly oscillating microwave field, which is non-selective for short pulses.  At  $B=0.22$ T the rotation speed ratio is $|\eta_{12\to11}|/|\eta_{9\to8}|=5/4$. A  $5\pi$ rotation
 of the nuclear qubit corresponds to a $4\pi$ rotation of the unwanted $|9\rangle\to |8\rangle$ transition. In {\bf (d)} we use a RH circularly polarised microwave field, which selects for the desired conditional nuclear qubit rotation, and is much more efficient than the linearly polarised case. } 
\label{Fig5} 
\end{figure}

For example:   Consider  the initial state $|\psi\rangle=\frac{1}{\sqrt{2}}\left( |1_e1_n\rangle + |0_e1_n\rangle \right)$. If we wanted to perform a CNOT gate on this state, with the electron qubit as the control, we might choose to use the transition frequency $\omega^{12\leftrightarrow11}$ as 
dictated by Table \ref{table1}. However, this frequency is only a few MHz different from that of $\omega^{9\leftrightarrow8}$, and a short pulse of $\sim$ 50 ns  would  also drive the transition between states $|9\rangle$ and $|8\rangle$, and thereby effect an unwanted operation on our qubits. There are two strategies to overcome this complication: 

\begin{enumerate}
\item 
by tuning the microwave frequency to be exactly between the wanted and unwanted transition frequencies and assuming  a square pulse, we ensure they are both affected by the same pulse power $\omega_1'=\omega_1\sinc(T\delta \omega_0)$, where T is the pulse duration, such that the only variable affecting the two transition rates would be the mixing factor $\eta$. Near the $m=-4$ cancellation resonance at 
$B=0.22$ T, we get $|\eta_{12\to11}|/|\eta_{9\to8}|=5/4$. This ensures that at time $t=10\pi/(\omega_1'|\eta_{12\to11}|)$, we have performed the operation
\begin{equation}
 \frac{1}{\sqrt{2}}\left(|1_e1_n\rangle+|0_e1_n\rangle\right) \to  \frac{1}{\sqrt{2}}\left(-i|1_e0_n\rangle+|0_e1_n\rangle\right).
\end{equation}

\noindent This is shown numerically in Fig.\ref{Fig5}(c).  

\item
The transition $|12\rangle\to|11\rangle$ utilises the RH component of the microwave field, whereas the $|9\rangle\to|8\rangle$ transition uses the LH one. By generating a RH circularly polarised microwave field, we would be able to select for the desired transition, as shown in Fig.\ref{Fig5}(d).

\end{enumerate}

Obviously, scheme $2$ is preferable as it requires much shorter times to carry out our quantum gates. This scheme can  be used in selecting for   one of the transitions $|\pm,m\rangle \leftrightarrow |\mp,m-1\rangle$, which differ in frequency by $2\delta\omega_0$, and, similarly, for transitions $|\pm,m\rangle \leftrightarrow |\pm,m-1\rangle$.  
 
\subsection{Scaling with controllable Heisenberg interaction}

So far we have only described how to perform two-qubit gates in a single site of Si:Bi (or any other nuclear-electronic system obeying the same Hamiltonian and with a large enough hyperfine exchange term). This is very limited, however, and we need to be able to scale the system so as to incorporate arbitrarily large numbers of qubits. With Si:Bi, the possibility does exist to further utilise the 20 dimensional Hilbert space, which would provide a maximum of four qubits. This is, however, not scalable, and we will still be limited to just four qubits as we cannot create more energy levels within the single-site Si:Bi system. The only  feasible option that remains, is to have spatially separated Si:Bi centers, between which we can establish an interaction. The original Kane proposal \cite{Kane} envisaged a Heisenberg interaction between nearest neighbour electrons which could be controlled via changing the electrostatic potential barrier between the donor sites, that, in turn, alters the degree of electron wavefunction overlap. This interaction has the effective form:

\begin{equation}
\hat H_{int}=J\hat {\bf S}^i.\hat{\bf S}^{i+1}.
\end{equation}          

\noindent The same interaction could be achieved indirectly by modulating the Rydberg state of a control dopant placed in between the two qubits \cite{stoneham}. As shown by \cite{DiVincenzoSmolin1999} such an interaction can be used to produce a $\sqrt{\text{SWAP}}$ gate:

\begin{equation}
\sqrt{\text{SWAP}}=e^{i \frac{\pi}{8}}e^{-i\frac{\hat H_{int}}{J}\frac{\pi}{2}}.
\end{equation}

\noindent Using two such gates, together with single-qubit unitaries, we can establish a CZ gate between the electrons:
\begin{equation}
CZ^{12} = e^{i \frac{\pi}{2}}\left(e^{-i\hat \sigma^1_z\frac{\pi}{4}}\otimes e^{i\hat\sigma_z^2\frac{\pi}{4}}\right)\sqrt{\text{SWAP}}e^{-i\hat \sigma_z^1\frac{\pi}{2}}\sqrt{\text{SWAP}} 
\end{equation}

\noindent As stated in previous sections, we cannot perform rotations about the $z$ axis of the Bloch sphere directly, but we can use our EPR pulses about the $x$ and $y$ axes to produce the required single-qubit unitaries. 

\begin{equation}
e^{-i\hat \sigma^1_z\frac{\pi}{4}}=e^{i \pi}e^{-i\hat \sigma^1_y\frac{3\pi}{4}}e^{-i\hat \sigma^1_x\frac{\pi}{4}}e^{-i\hat \sigma^1_y\frac{\pi}{4}}.
\end{equation}

The CZ gate can be turned into a CNOT gate by simply applying a Hadamard on the target qubit before and after the application of the CZ. Such an interaction affects the electron spin basis states, and not the adiabatic basis states to which we have designated our logical qubits. Therefore, we must apply our electronic two-qubit gates in the high-field limit where mixing is suppressed, and where there is a high fidelity between the adiabatic basis and spin basis. A consequence of this is that the energy difference between different eigenstates will be very large and, as is well known  \cite{AndersonIsing}, if $|E_i -E_{i+1}| \gg J$, the Heisenberg interaction between the two effectively becomes an Ising interaction $J\hat S_z^1 \otimes\hat S_z^2$. To use the above scheme of producing entangling two-qubit gates between all four eigenstates in each of the two adjacent sites, we would have to establish a very strong $J$. 

Alternatively, we can set $B_i$ and $B_{i+1}$ to be sufficiently different, and $J$ small enough, such that we only get an Ising interaction between all relevant eigenstates. It is in fact easier to produce the CZ and CNOT gates with  an Ising interaction, as it only requires one exchange operation and not two as in the case of the Heisenberg interaction \cite{XYtwoqubit,Makhlin}.        

\begin{equation}
CZ^{12} = e^{-i \frac{\pi}{4}}\left(e^{i\hat \sigma_z^1\frac{\pi}{4}}\otimes e^{i \hat \sigma_z^2 \frac{\pi}{4}}\right)e^{-i\hat S_z^1 \otimes \hat S_z^2\pi}.
\end{equation}

\section{Decoherence from temporal magnetic field fluctuations}\label{section4}

For practical quantum information processing in silicon, the substance would need to be purified so as not to contain any $^{29}$Si such that no decoherence would result due to spin-bath dynamics. Temperatures would also be maintained at low levels in order to minimise the phonon-bath induced decoherence. Here, we wish to employ a phenomenological model of  decoherence for nuclear-electronic systems, resulting only from stochastic magnetic-field fluctuations. Taking the Hamiltonian from Eq.\eqref{HAM} and adding to it a perturbative term involving independent temporal magnetic-field fluctuations in all three spatial dimensions (with the usual association of $1=x,2=y,3=z$), all of which take a Gaussian distribution with mean 0 and variance $\alpha_n^2$, gives in the interaction picture:

\begin{equation}
\tilde H(t) = \sum_{n=1}^3\omega_n(t)\tilde S_n(t) +\omega_n(t)\delta\tilde I_n(t)
\end{equation}     

\noindent where $\{S_n\}$ and $\{I_n\}$ are the $n$-axis electron and nuclear spin operators respectively, and $\omega(t)$ is the electron Zeeman frequency at time $t$. As before, $\delta$ represents the ratio of the nuclear to electronic Zeeman frequencies, and because it is small we may  ignore the nuclear term. We then follow the standard procedure of deriving a Born-Markov master equation \cite{Breuer}:

\begin{eqnarray}
&&\frac{d}{dt}\langle\rho(t)\rangle = i\left[\langle\rho(t)\rangle,\hat H_0  \right]\nonumber  +\alpha_n^2\sum_{n=1}^3\sum_{\Omega} e^{-\chi_n \Omega^2} \times \nonumber \\
&&\left( \hat  S^\dagger_n(\Omega)\langle\rho(t)\rangle\hat  S_n(\Omega)-\frac{1}{2}\left[ \langle\rho(t)\rangle,\hat  S^\dagger_n(\Omega) \hat S_n(\Omega)\right]_+  \right)
  \label{masterequation}\end{eqnarray}

\noindent where $\hat S_n(\Omega)$ are the electron spin operators in the eigenbasis of $\hat H_0$,  $\Omega$ is the energy difference between two such eigenstates, and $\chi_n = (dB_n/dt)^{-1}$ is the inverse of the rate of change of the magnetic field strength in direction $n$. $\langle \rho(t)\rangle$ is defined as the  density operator of the coupled nuclear-electronic system, averaged over either an ensemble of such systems, or repeated experiments on a single system.  Further details for the derivation of Eq.\eqref{masterequation} can be found in Appendix \ref{masterequationderivation}.

The rate term $e^{-\chi_n\Omega^2}$ imposes the results of the adiabatic theorem into our master equation. The quantitative condition for adiabatic evolution is often cited as \cite{kwek}

\begin{equation}
\left |\frac{\langle\phi_i|\dot H(t)|\phi_j\rangle}{{\Omega^{i\leftrightarrow j}}^2}\right| \ll 1 .
\end{equation}

\noindent For the model described here, this translates to 

\begin{equation}
\left |\langle\phi_i|\hat S_n|\phi_j\rangle\frac{ \frac{dB_n}{d t}}{{\Omega^{i\leftrightarrow j}}^2}\right|=\left|\langle\phi_i| \hat S_n|\phi_j\rangle \right| \frac{1}{\chi_n {\Omega^{i\leftrightarrow j}}^2} \ll 1 
\end{equation} 

\noindent which means that, in the case of $\left|\langle\phi_i| \hat S_n|\phi_j\rangle \right| >0$,  if the magnetic field is fluctuating sufficiently slowly,  the probability of transition between eigenstates $|\psi\rangle$ and $|\phi\rangle$ becomes vanishingly small. 

Since we have made the rotating wave (or secular) approximation, and are only interested in the interaction picture dynamics of our system, we may drop the Hamiltonian commutator in Eq.\eqref{masterequation}, leaving only the dissipator term. For Si:Bi in the intermediate-field regime, this condition is satisfied by setting $\alpha^2/2\pi=9$ MHz, and we  use this value whenever we provide numerical calculations.  We are now equipped with the tools to address decoherence in our quantum system. We may model our  gates as ideal unitaries that can prepare some superposition $|\psi\rangle = a|e\rangle + b|g\rangle$ between the  adiabatic basis states $|e\rangle$ and $|g\rangle$, which is then decohered according to our noise model. The  dephasing and depolarising rates   are determined by applying our master equation and measuring the rate that the  observables 
\begin{equation}
\sqrt{\mathrm{tr}[\sigma_x \tilde\rho(t)]^2+\mathrm{tr}[\sigma_y\tilde\rho(t)]^2}
\end{equation}
  \noindent and
 \begin{equation}
 \mathrm{tr}[\sigma_z\tilde\rho(t)]
 \end{equation}
  \noindent decay respectively. In the cases that these decay as $e^{-\Gamma t}$, where $\Gamma$ is the decay rate, we may characterise the dephasing and depolarising times by  $T_2$ and $T_1$ respectively, which are the inverse of the decay rates. Such times are  measured   in EPR experiments \cite{Feherrelaxation,Orbach}. The Pauli matrices denoted here are in the eigen-basis of the reduced two-level system in question. We will study two types of noise; $Z$ noise and $X$ noise, so named due to  Gaussian magnetic-field fluctuations in the $z$-axis and $x$-axis respectively. Throughout this section, when an adiabatic state is indicated with $m$, it is implied that $|m| < (I + \frac{1}{2})$, and states where $m=\pm(I+\frac{1}{2})$ are explicitly designated.

\subsection{$Z$ noise}

\begin{figure}[!htb] 
\includegraphics[width= 3.5 in]{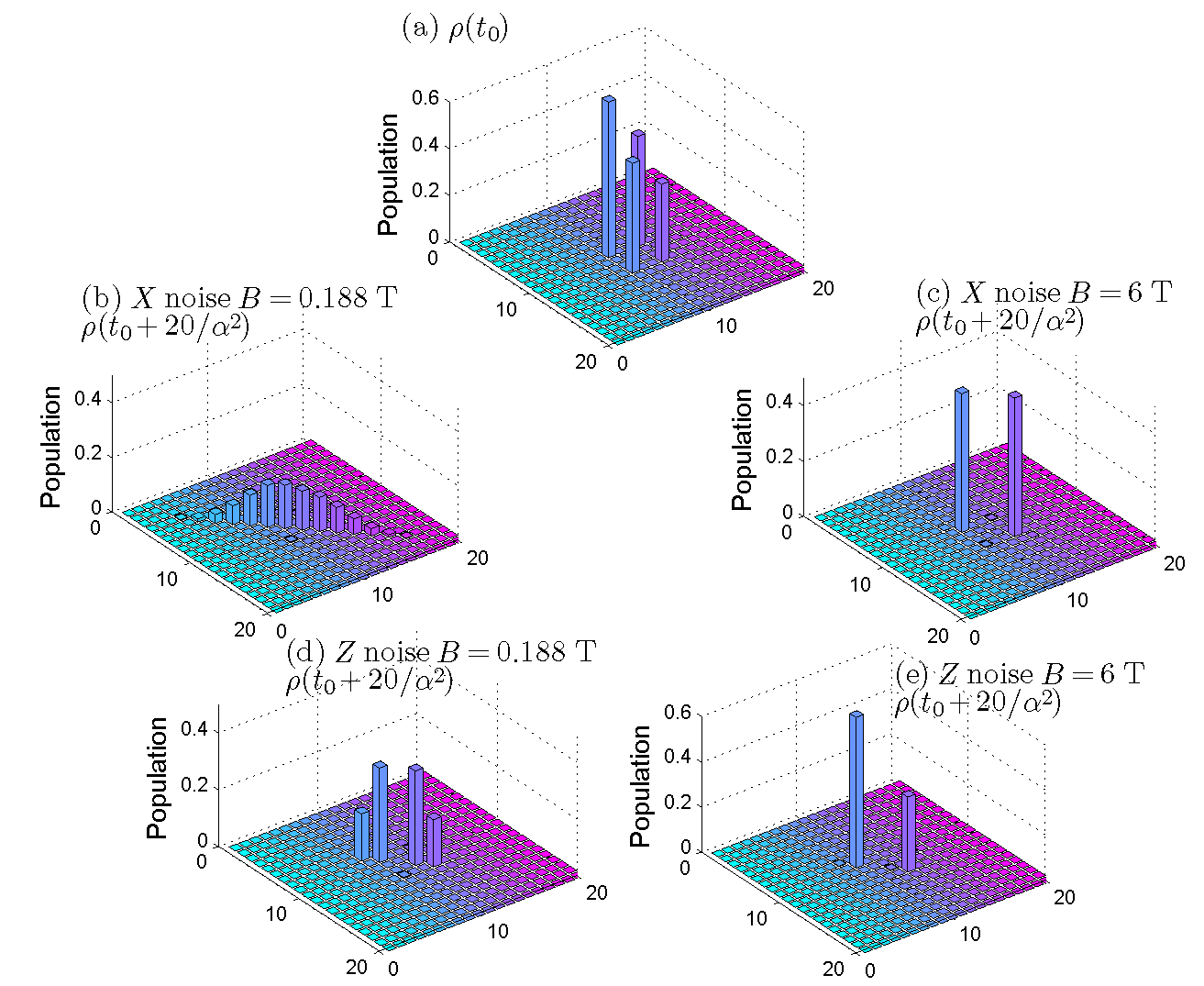} 
\caption{(Colour online) Shows the change in density operator elements of $|\psi\rangle=\frac{2}{\sqrt{3}}|9\rangle+\frac{1}{\sqrt{3}}|12\rangle$ after a period of $20/\alpha^2$, given diabatic Gaussian noise of variance $\alpha^2$, at two field regimes: the frequency minimum of $B=0.188$ T, and $B=6$ T . (a) Population of $\rho$ in the eigen-basis of $ \hat H_0$ at time $t_0$. (b) and (c) show the effect of $B$ on  $X$  noise. (b) Near the frequency minima, $X$ noise couples every part of the full Hilbert space, and  depolarises the full system, ultimately resulting in $\frac{1}{20}\mathds{1}$. (c) At $B=6$ T, $X$ noise decouples states $|+,m\rangle,|-,m-1\rangle$  from the rest of the Hilbert space, and effects a depolarising channel in that subspace only. (d) and (e) show the effect of $B$ on  $Z$ noise. Z noise conserves angular momentum and hence keeps to the four-dimensional Hilbert space of $m=-3, m-1 = -4$. (d) shows that  at the frequency minimum, $Z$ noise effects independent depolarising channels for each $m$ subspace. As a result, the population of states $|12\rangle$ and $|9\rangle$ equalise with those of states $|8\rangle$ and $|11\rangle$ respectively.(e) shows that at $B = 6$ T, $\langle+,m|\hat S_z|-,m\rangle \sim0$ and we simply get a dephasing channel for the $|+,m\rangle,|-,m-1\rangle$ subspace.      }   
\label{Fig6}
\end{figure}

 Given  $Z$ noise,  we may consider our system as a decoupled four-level system  with sub Hamiltonian:

\begin{equation}
H_{sub}=H_{m}^{2d}\oplus H_{m-1}^{2d}.
\end{equation}

\noindent  This is possible as there will be no transfer of population to other components of the Hilbert space. We may write  $\hat S_z$  in the adiabatic basis $\left\{|+,m\rangle,|-,m\rangle,|+,m-1\rangle,|-,m-1\rangle\right\}$:

\begin{equation}
\begin{pmatrix} \cos(\theta_m) & -\sin(\theta_m) & 0 & 0 \\
-\sin(\theta_m) & -\cos(\theta_m)& 0 & 0 \\
0 & 0 & \cos(\theta_{m-1}) & -\sin(\theta_{m-1}) \\
0 & 0 & -\sin(\theta_{m-1}) & -\cos(\theta_{m-1})
\end{pmatrix}\label{Znoiseadbasis}
\end{equation}

\noindent Substituting this into Eq.\eqref{masterequation} and considering it in the interaction picture gives:

\begin{eqnarray}
\frac{d}{dt}\langle\tilde\rho(t)\rangle &=&\sum_{n=m-1}^m   
 \frac{\alpha^2}{4}\cos^2(\theta_n) [ \hat \sigma_z^n\langle\tilde\rho(t)\rangle\hat \sigma_z^n- \langle\tilde\rho(t)\rangle  ]  \nonumber \\
 && + \frac{\alpha^2}{4}e^{-\chi \Omega^2}\sin^2(\theta_n) \times \nonumber \\ &&[  |+,n\rangle\langle -,n|\langle\tilde\rho(t)\rangle|-,n\rangle\langle +,n| - \nonumber \\&&     |+,n\rangle\langle -,n||-,n\rangle\langle +,n|\langle\tilde\rho(t)\rangle + \mathrm{H.c.}]. 
\label{masterequationznoise}
\end{eqnarray}

In the high-field limit  our EPR local unitaries can only  create superpositions $a|+,m\rangle + b|-,m-1\rangle$. As the noise operator takes the form $\hat \sigma_z^m\oplus\hat \sigma_z^{m-1}$ in this regime, this superposition  may be considered to exist  as a decoupled two-level system. We may therefore  solve Eq.\eqref{masterequationznoise}  (for the two-level subspace in question) analytically:

\begin{equation}
e^{\mathcal{L}t}\tilde\rho(t_0)=\frac{1}{2}\left(1 +e^{-\frac{t}{T_2}}\right)\tilde\rho(t_0)+ \frac{1}{2} \left( 1 -e^{-\frac{t}{T_2}}\right)\hat \sigma_z\tilde\rho(t_0)\hat\sigma_z.\label{dephasingchannel}
\end{equation}   

\noindent where $\mathcal{L}$ is the Liouville superoperator, whose action on $\rho$ is given by Eq.\eqref{masterequationznoise}. This is simply the dephasing channel for a spin $1/2$ particle \cite{NielsenChuang}:
\begin{equation}
\mathcal{E}(t)\circ\rho=(1-\lambda(t))\rho +\lambda(t)\hat\sigma_z\rho\hat\sigma_z
\end{equation}
 with probability $\lambda(t)$ of performing a $\hat\sigma_z$ operation under conjugation. Here, $\lambda(t)=\frac { 1 -e^{-\frac{t}{T2}}}{2}$ with a $T_2$ time of  $2/\alpha^2$. This is illustrated by Fig.\ref{Fig6}(e), where only the off-diagonal elements of $\rho(t_0)$, as shown in Fig.\ref{Fig6}(a), decay.  
At low fields however, $\langle +,m|\hat S_z|-,m\rangle >0$ and we cannot ignore the exchange term in Eq.\eqref{masterequationznoise}. In this case, we may use the four-dimensional Bloch vector representation of our density operator:  

\begin{equation}
\rho(t)=\frac{1}{4}\left(\sum_{i,j=0}^3 n_{ij}(t) \hat\sigma_i\otimes\hat\sigma_j \right) \text{,}\ n_{00}(t)=1.
\end{equation}

\noindent It is possible to map the dynamics of the density operator to that of the Bloch vector \cite{Nazir} as\begin{equation}
\frac{d\bold{\underline{n}}(t)}{dt} = \mathcal{L}\bold{\underline{n}}(t).
\end{equation}

\noindent For Si:Bi the 16 simultaneous differential equations can be solved to obtain analytic expressions for the dephasing and depolarising rates. Alternatively, by decomposing $\bold{\underline{n}}(t)$ in the eigenbasis of $\mathcal{L}$, denoted $\bold{\underline{n}}_l$ with generally complex eigenvalues $\lambda_l$, we may represent  the dynamics of the Bloch vector as

\begin{equation}
\bold{\underline{n}}(t)=\sum_{l=0}^{15} c_l \bold{\underline{n}}_le^{t\lambda_l}.
\end{equation}

\noindent where $c_l$ are determined by the initial conditions. It is the real component of the eigenvalues which leads to decay in population of the eigenstate. The infinite time state is therefore a superposition of eigenstates $\bold{\underline{n}}_l$ such that $Re(\lambda_l)=0$.

\subsubsection{Adiabatic $Z$ noise}

Here we may set $\chi\to \infty \implies e^{-\chi\Omega^2}=0$ for $\Omega^2 > 0$. There will be no depolarisation in this case, and we may only have pure dephasing. For superpositions of type $a|\pm,m\rangle + b|\mp,m-1\rangle$, the dephasing rate, parameterised as the decay of the off-diagonal elements of the subspace in question, is given by
\begin{equation}
\frac{1}{T_2}= \frac{\alpha^2}{8}[\cos(\theta_m)+\cos(\theta_{m-1})]^2.
 \end{equation}
 
  \noindent When $\cos(\theta_m) = -\cos(\theta_{m-1})$, which is satisfied at the frequency minima, dephasing due to $\hat S_z$ is completely removed. 

For superpositions of type $a|\pm,m\rangle + b|\pm,m-1\rangle$, the dephasing rate is given by 
\begin{equation}
\frac{1}{T_2} =\frac{\alpha^2}{8}[\cos(\theta_m)-\cos(\theta_{m-1})]^2.
 \end{equation}
  \noindent Here there are two regions where the  $\hat S_z$ caused dephasing is removed; when $\cos(\theta_m) = \cos(\theta_{m-1})$ which occurs at the frequency maxima, and at the high-field limit where $\cos(\theta_m)=\cos(\theta_{m-1})=1 \ \forall \ m$, rendering such transitions as only NMR-allowed.  

For superpositions of type $a|\pm,\pm(I+\frac{1}{2})\rangle+b|\pm,m\rangle$ the  dephasing rate is given by 
\begin{equation}
\frac{1}{T_2}=\frac{\alpha^2}{2}\sin\left(\frac{\theta_m}{2}\right)^4
\end{equation} 

\noindent which reaches its minimal value of 0 as $B \to \infty$, whereas for superpositions of type $a|\pm,\pm(I+\frac{1}{2})\rangle+b|\mp,m\rangle$ it is given by
\begin{equation}
\frac{1}{T_2}=\frac{\alpha^2}{2}\cos\left(\frac{\theta_m}{2}\right)^4
\end{equation}

\noindent which reaches its minimal value (which is generally greater than 0) at $B=0$ T. The steady state solution for adiabatic $Z$ noise  is given by 

\begin{equation}
\bold{\underline{n}}(\infty)=\mathds{1}\otimes\mathds{1}+c_1\mathds{1}\otimes\hat \sigma_z+c_2\hat\sigma_z\otimes\mathds{1}+c_3\hat\sigma_z\otimes\hat\sigma_z. \end{equation}

\subsubsection{Diabatic $Z$ noise}

Here we may set $\chi \to 0 \implies e^{-\chi\Omega^2}=1 \  \ \forall \ \Omega^2$. Solving the Bloch vector differential equations yields analytic expressions for the dephasing rates. For an initial superposition of $a|\pm,m\rangle +b |\mp,m-1\rangle$ this gives:
\begin{equation}
\frac{1}{T_2}=\frac{\alpha^2}{4}\left(\cos(\theta_m)\cos(\theta_{m-1})+1\right) \label{diabaticzalloweddephasing}\end{equation}

\noindent and for $a|\pm,m\rangle + b|\pm,m-1\rangle:$
\begin{equation}
\frac{1}{T_2} = \frac{\alpha^2}{4}\left(\cos(\theta_m)\cos(\theta_{m-1})-1\right) \label{diabaticzforbiddendephasing}.\end{equation}

\noindent Eq.\eqref{diabaticzalloweddephasing}  reaches a minimum value (hence giving the longest $T_2$ time)  when $\cos(\theta_m)=-\cos(\theta_{m-1})$, i.e., at the frequency minima. Unlike  the adiabatic $Z$ noise case, this value does not reach 0, but rather reaches approximately half its maximal value at the high-field regime.  Conversely,  Eq.\eqref{diabaticzforbiddendephasing} reaches its maximal value when $\cos(\theta_m)=-\cos(\theta_{m-1})$, attaining approximately the same value as  for  Eq.\eqref{diabaticzalloweddephasing} at this regime. Note that unlike  the adiabatic case, there is no decoherence minimum at the frequency maxima; the decay rate simply vanishes as $B \to \infty$. 

Because the exchange terms in Eq.\eqref{masterequationznoise} contribute to the dynamics for  diabatic $Z$ noise, there will also be depolarising noise in each $m$ subspace, equalising the population in states $|\pm,m\rangle$ . Fig.\ref{Fig6}(d) shows the effect of this depolarisation at the frequency minima. The depolarisation rate of each $m$ subspace   is given by:

\begin{equation}
\frac{1}{T_1}  =\frac{\alpha^2}{2}\sin(\theta_m)^2,
\end{equation}

\noindent which vanishes as $B \to \infty $, and maximises at the avoided crossing cancellation resonance. Given any superposition $\sqrt{P_g}|g\rangle +e^{i\phi}\sqrt{P_e}|e\rangle$,  with states $|g\rangle$ and $|e\rangle$ each existing in a different $m$ subspace, such that $\mathrm{tr}\left[\hat H_0(|e \rangle\langle e|-|g\rangle\langle g|)\right] >0$, the depolarisation is given by

\begin{equation}
\mathrm{tr}[\sigma_z\tilde\rho(t)] =\frac{1}{2}P_e\left(1 +e^{-t/T_1^e}\right) - \frac{1}{2}P_g\left(1 +e^{-t/T_1^g}\right)  \label{effectivedepolarisation}
\end{equation}

\noindent where $1/T_1^{g/e}$ is the depolarisation rate in the $m$ subspace for $|g\rangle$ and $|e\rangle$, respectively. 
The steady state solution for diabatic $Z$ noise given such superpositions is given by 

\begin{equation}
\bold{\underline{n}}(\infty)=\mathds{1}\otimes\mathds{1}+c_1\hat\sigma_z\otimes\mathds{1} \end{equation}

\noindent where $c_1 \in[1,-1]$.
For superpositions of type $a|\pm,\pm(I+\frac{1}{2})\rangle+b|\pm,m\rangle$ the dephasing rate is given by 
\begin{equation}
\frac{1}{T_2}=\frac{\alpha^2}{4}[1 - \cos(\theta_m)]
\end{equation}

\noindent and for superpositions of type $a|\pm,\pm(I+\frac{1}{2})\rangle+b|\mp,m\rangle$ the dephasing rate is given by 
\begin{equation}
\frac{1}{T_2}=\frac{\alpha^2}{4}[1 + \cos(\theta_m)].
\end{equation}

\noindent The depolarisation rate can be calculated as in the previous case, using Eq.\eqref{effectivedepolarisation}, and noting that one of $T_1^{g/e}$ is equal to $\infty$.
The steady state solution for diabatic $Z$ noise for such superpositions is given by 

\begin{equation}
\bold{\underline{n}}(\infty)=\mathds{1}\otimes\mathds{1}+c_1\hat\sigma_z\otimes\mathds{1} + c_2\mathds{1}\otimes\hat \sigma_z. \end{equation}
 
\begin{figure}[!htb] 
\includegraphics[width= 3.5 in]{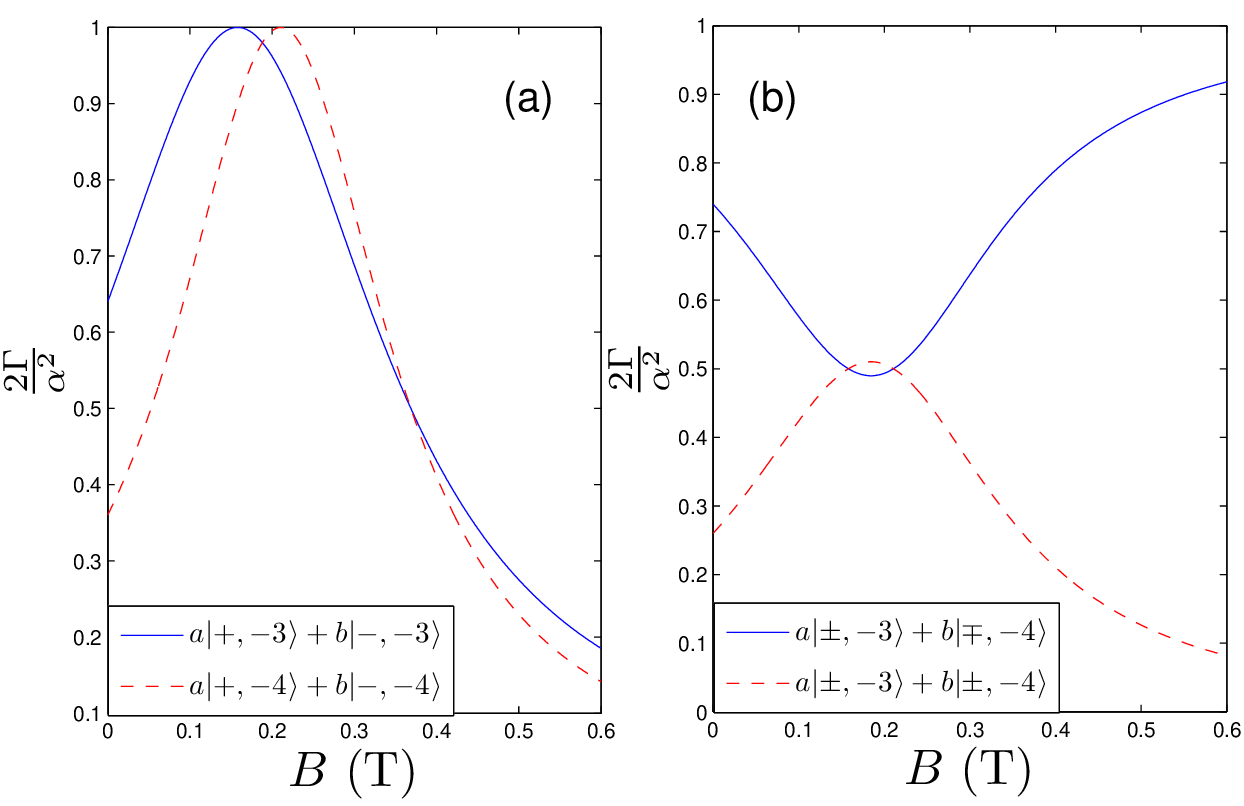} 
\caption{(Colour online) The exponential decay rate given by $\Gamma$ in units of $\alpha^2/2$ for diabatic $Z$ noise driven (a) depolarisation and (b) dephasing in Si:Bi. This is done in the four-dimensional subspace of $m=-3,m-1=-4$. (a) shows that in each subspace, the depolarisation rate maximises when $\theta_m=\pi/2$, or the avoided crossing cancellation resonances. (b) shows that at the high-field limit, the dephasing rate of $a|\pm,m\rangle + b|\mp,m-1\rangle$ is maximal, whilst that of $a|\pm,m\rangle+b|\pm,m-1\rangle$ becomes vanishingly small. It should be noted, however, that the $a|-,m\rangle + b|+,m-1\rangle$ superposition cannot be made by either EPR or NMR in the high-field limit. These rates both approximately reach the value of 1/2 at the frequency minima.    }   
\label{Fig12} 
\end{figure}

\noindent For diabatic $Z$ noise, Fig.\ref{Fig12} shows the analytical depolarisation and dephasing rates for subspace $m=-3,m-1=-4$ in Si:Bi. Fig.\ref{Fig8}(a) shows the numerically calculated $T_2$ times for all EPR lines depicted in Fig.\ref{Fig2}(b), whilst  Fig.\ref{Fig9}(a) shows the numerically calculated  depolarising times $T_1$ in units of $2/\alpha^2$ for each $m$-subspace. 

\subsection{$X$ noise}

$X$ noise is less trivial, as it  couples all components of the Hilbert space so we cannot consider a sub-Hamiltonian in isolation. Solving the resulting 400 Bloch equations (for Si:Bi) would be unfeasible, so only numerical calculations are given here. Furthermore, the adiabatic condition must be violated for $X$ noise to have any effect, as there are no $\hat S_x(\Omega=0)$ terms in Eq.\eqref{masterequation}.  In the high-field limit the $X$ noise operator  will  take the form of $\hat\sigma_x\otimes\mathds{1}$ in the basis $\{|+,m\rangle  ,  |-,m-1\rangle \}$ as well as $\{|-,m\rangle  ,  |+,m-1\rangle \}$. At such fields, as shown in Fig.\ref{Fig6}(c), an arbitrary superposition of $a|\pm,m\rangle +b|\mp,m-1\rangle$ suffers a two-level system depolarising channel. At low fields, however, the dissipation is not contained within the $m,m-1$ subspace, and as indicated by Fig.\ref{Fig6}(b) the system eventually decays to $\frac{1}{d}\mathds{1}$.     

\begin{figure}[!htb] 
\includegraphics[width= 3.5 in]{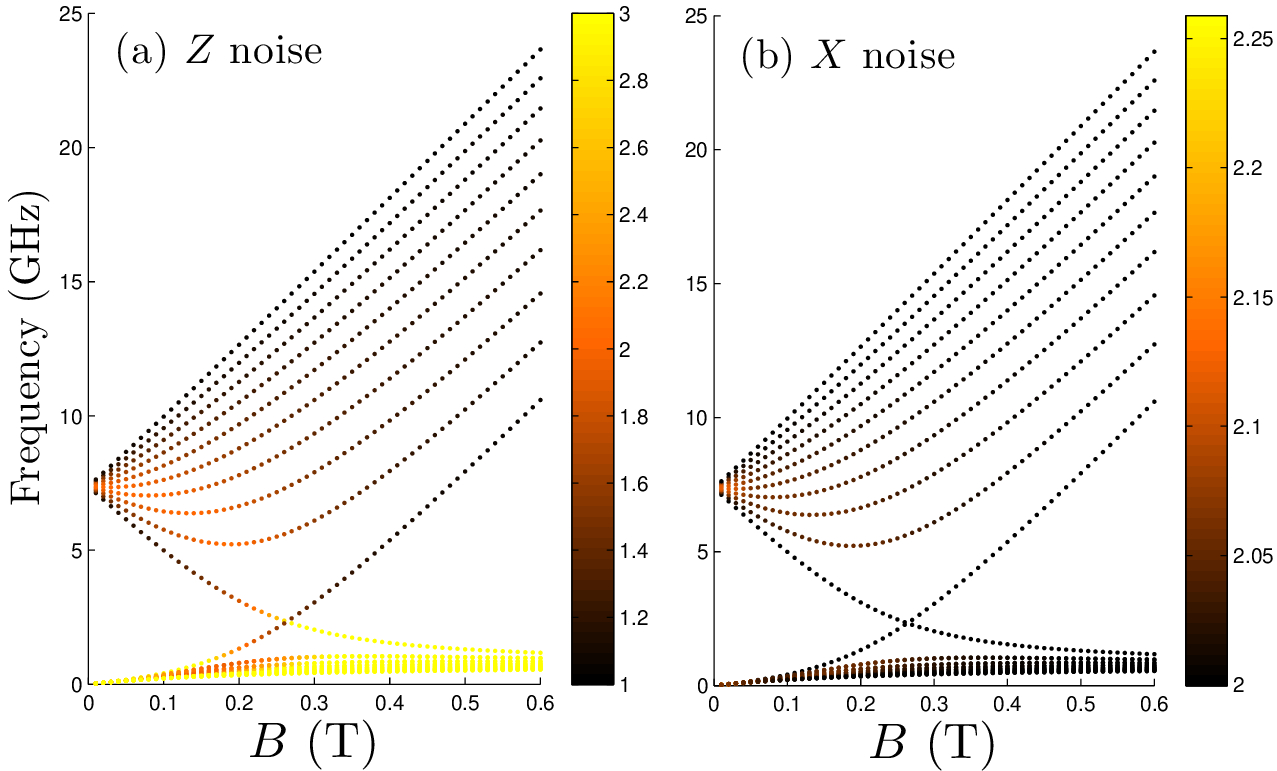} 
\caption{ (Colour online) Simulated dephasing times in units of $2/\alpha^2$ for diabatic (a) $Z$  and (b) $X$ noises, calculated with $\alpha^2/2\pi= 9$ MHz.  In (a) the superpositions $a|\pm,m\rangle+b|\mp,m-1\rangle$ have $T_2$ times of $2/\alpha^2$ at $B \gtrsim 0.6$ T, and approximately $4/\alpha^2$ at the frequency minima. Superpositions $a|\pm,m\rangle+b|\pm,m-1\rangle$ also have $T_2$ times of $4/\alpha^2$ at the frequency minima. However, as $B$ increases, these become NMR transitions and will have $T_2$ times of $2/(\alpha^2\delta)$. The colour bar has been truncated after three to aid visibility but the maximum value reaches as high as $\sim 100$. In (b), the $T_2$ time does not vary by much, and reaches its maximal points at fields less than the frequency minima.   }   
\label{Fig8} 
\end{figure}

For $X$ noise, all dephasing is a result of the depolarising noise that is effected by the $X$ noise operator, and as shown in Fig.\ref{Fig8}(b), at $B > 0.6 $ T the dephasing time is $4/\alpha^2$ for all  transitions. This value increases only slightly at magnetic fields smaller than the frequency minima for transitions involving $m < 0$.

\begin{figure}[!htb] 
\includegraphics[width= 3.5 in]{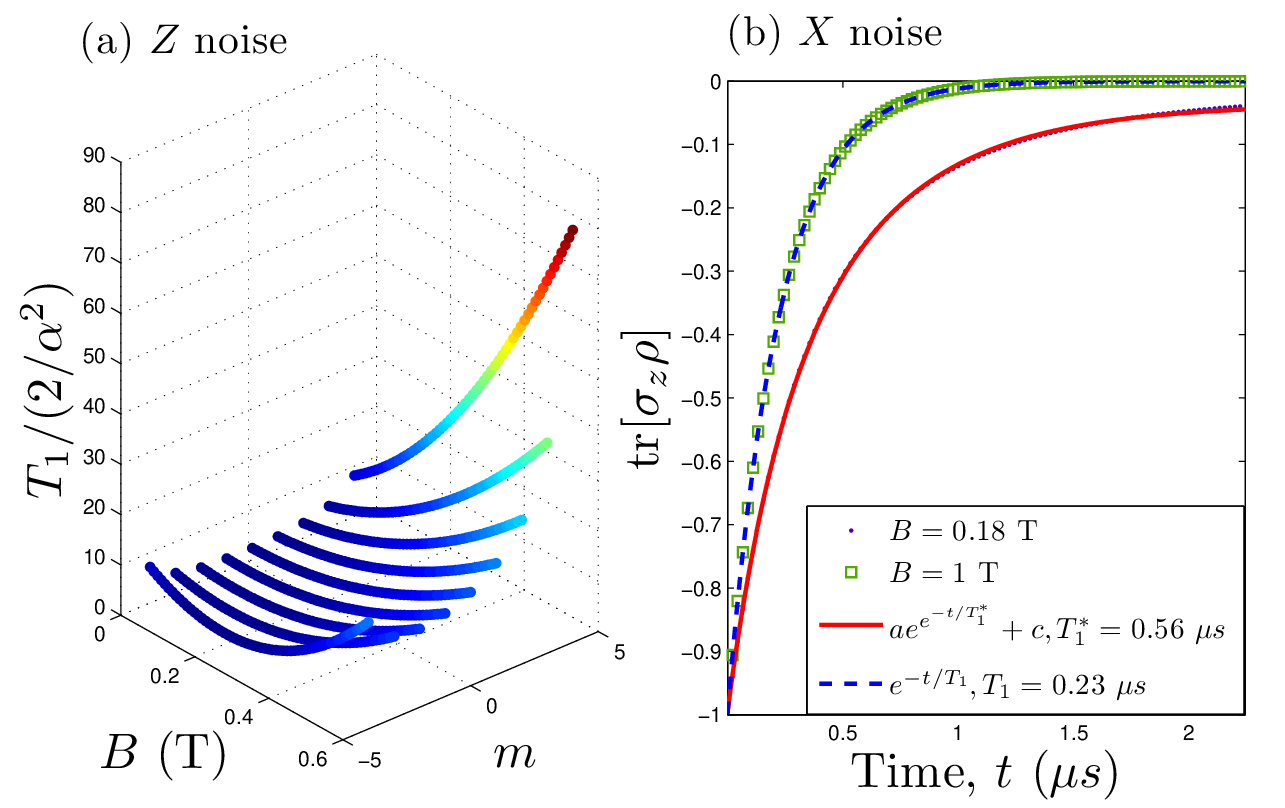} 
\caption{ (Colour online) Simulated depolarising times for diabatic $Z$ and $X$ noise with $\alpha^2/2\pi = 9$ MHz. (a) Given $Z$ noise, the decay of $\mathrm{tr}[\sigma_z\tilde\rho(t)]$ within each $m$ subspace is always exponential. For $m \leq 0$ the $T_1$ time reaches a minimum at the avoided crossing cancellation resonances. The $T_1$ time for subspace $m$ and $m-1$ become identical at the frequency minima (b) Given $X$ noise, the decay in the $\{|+,m\rangle \ |-,m-1\rangle\}$ and $\{|-,m\rangle \ |+,m-1\rangle\}$ subspaces follows an exponential curve at high magnetic fields, but at  low magnetic fields such as the frequency minima, it follows a double exponential fit.     }   
\label{Fig9} 
\end{figure}

Fig.\ref{Fig9}(b) shows the different forms of depolarising rates for $X$ noise. At high magnetic fields, where the only non-vanishing matrix elements of the $X$ noise operator are $\langle \pm,m|\hat S_x|\mp,m-1\rangle$, the depolarising noise follows an exponential decay. Under intermediate magnetic fields however, the dissipation follows a more complicated mechanism and the decay is better explained by a double exponential fit. 

\section{Conclusions}\label{section5}
A coupled nuclear-electronic spin system with large A will have its eigenstates as superpositions of the $z$-axis spin basis states at appreciably large magnetic fields, which we call the intermediate-field regime. This will allow for performing EPR transitions between eigenstates that, at high-field, are EPR-forbidden, and would require NMR pulses which are  orders of magnitude slower. We have shown that this allows for two-qubit universal quantum computation to be performed with only the use of EPR pulses. Si:P has $A/2\pi=117.5$ MHz, so it will be in the intermediate-field regime when $B\sim0.02$ T. At such a low field the transition frequencies are of the order $\lesssim 0.5$ GHz. With Si:Bi on the other hand, with $A/2\pi=1.4754$ GHz, the intermediate-field condition is satisfied when $B \sim 0.5$ T  and the transition frequencies are of order $\lesssim 10$ GHZ. For current EPR technology, operation in the intermediate-field regime is easier to carry out on Si:Bi. Indeed, this has been recently demonstrated experimentally in \cite{Zurich}. 

For a nuclear-electronic spin system with $I\geq 1$,  cancellation resonances can be seen at non vanishing magnetic fields; Si:P has only one cancellation resonance at $B\simeq0$ T, whereas Si:Bi has a series of cancellation resonances at $B \lesssim0.3$ T. Furthermore, interesting effects such as decoherence reduction, associated with $df/dB=0$ points, occur between eigenstates belonging to two different subspaces that have a cancellation resonance. As a result, Si:P with $I=1/2$ does not have any $df/dB=0$ regions and hence no decoherence reduction points, whereas Si:Bi with $I = 9/2$ has several. The combination of fast EPR quantum gates and decoherence reduction makes Si:Bi an attractive system for quantum information processing.


\acknowledgements

M. Hamed Mohammady acknowledges an EPSRC studentship, and Ahsan Nazir  thanks Imperial College London and the EPSRC for financial support. Gavin Morley is supported by an 1851 Research Fellowship and the EPSRC COMPASSS grant.   The authors would like to thank Dara P. S. McCutcheon for the insightful discussions that helped the development of the decoherence theory in this paper.

\appendix

\section{Selective rotations} \label{unitarydynamics} 

Consider the coupled nuclear-electronic spin system in the eigenbasis of the Hamiltonian $\hat H_0$ given by Eq.\eqref{HAM}. The electron spin operators in this basis are given by the unitary transformation    

\begin{equation}
\hat S_x' = V^\dagger\hat S_xV \ \ \ \ \hat S_y' = V^\dagger\hat S_yV
\end{equation}

\

\noindent where $V$ is a matrix whose $i^{th}$ column is the $i^{th}$ eigenvector of $\hat H_0$. We want to be able to isolate two eigenstates of this Hamiltonian, and perform unitary dynamics in that subspace.  Tracing out all  eigenvectors other than $|e\rangle$ and $|g\rangle$  gives 

\begin{eqnarray}
( \hat S_x')^{eg} &=&  \frac{1}{2}\begin{pmatrix}0 & |e\rangle\langle e|(\hat S_++\hat S_-)|g\rangle\langle g| \\
|g\rangle\langle g|(\hat S_+ + \hat S_-)|e\rangle\langle e| & 0 \\
\end{pmatrix}\nonumber \\
&=&  \frac{\eta}{2} \hat \sigma_x \nonumber \\
(\hat S_y')^{eg} &=&  \frac{i}{2}\begin{pmatrix}0 & |e\rangle\langle e|(\hat S_- -\hat S_+)|g\rangle\langle g| \\
|g\rangle\langle g|(\hat S_- - \hat S_+)|e\rangle\langle e| & 0 \\
\end{pmatrix} \nonumber \\
&=& \mathrm{sign}_y \frac{\eta}{2} \hat \sigma_y
\end{eqnarray}

\noindent where $\eta = \langle e|\hat S_x|g \rangle$ is a measure of basis state mixing and $\mathrm{sign}_y=\langle e|\hat S_z + \hat I_z |e\rangle -\langle g |\hat S_z + \hat I_z|g \rangle \in\{1,-1\}$. As the absolute energies given by the eigenvalues are meaningless physically, we can re-scale the eigenvalues of $\hat H_0$ by adding to it an identity term   $-\frac{(\lambda_e +\lambda_g)}{2}\mathds{1}$ such that $\lambda_e$ and $\lambda_g$ are the eigenvalues of eigenvectors $|e\rangle$ and $|g\rangle$ respectively, where $\lambda_e > \lambda_g$. This gives:
\begin{equation}
\hat H_0 \mapsto \hat H_0^{\varsigma}= \frac{\Omega_0}{2}\hat \sigma_z^{eg} \oplus \hat H_0^{\text{rem}}   
\end{equation}

\noindent such that $\Omega_0 = |\lambda_e -\lambda_g|$, and $\sigma_z^{eg}$  exists in the $\{|e\rangle,|g\rangle\}$ subspace. Given a perturbative Hamiltonian of the form in Eq.\eqref{Vxyp}, and assuming that all EPR-allowed transition frequencies are unique, we may solve the Liouville-von Neumann equation for the two-level subsystem in the rotating frame of $\hat H_0^\varsigma$, while making the rotating wave approximation: 

\begin{widetext}

\begin{eqnarray}
\frac{d}{dt}\tilde\rho(t)&=&i\frac{\omega_1}{2}\left[\tilde\rho(t),e^{itH_0^{\varsigma}} \left\{   \left(\cos(\Omega_0 t)(\hat S_{x/y}')^{eg} + \sin(\Omega_0 t)(\hat S_{y/x}')^{eg}  \right) +\left(\cos(\Omega_0 t)(\hat S_{x/y}')^{eg} - \sin(\Omega_0 t)(\hat S_{y/x}')^{eg}   \right)  \right\}e^{-itH_0^{\varsigma}}  \right] \nonumber \\
 &=& i\frac{\omega_1\eta}{4}\left[\tilde \rho(t),\hat \sigma_{x/y} \right]  + i\frac{\omega_1\eta}{4}\left[\tilde \rho(t), \cos(2\Omega_0t)\hat \sigma_{x/y}  - \sin(2\Omega_0t)\hat \sigma_{y/x}\right] \nonumber \\ &\approx& i\frac{\omega_1\eta}{4}\left[\tilde \rho(t),\hat \sigma_{x/y} \right] \ \ \text{if \ } \omega_1 \ll \Omega_0. \label{interactionpictureresonanceevolution}
\end{eqnarray}

\end{widetext}

There are two possible regimes for the dynamics of this system. Those for which $\mathrm{sign}_y$, is positive(negative), which occurs when increasing the energy of the system corresponds to an increase(decrease) in $z$-axis total magnetisation $m$. In the case that $\mathrm{sign}_y=1$, the circular polarisation needed to achieve resonance with the Hamiltonian is of the form $\cos(\Omega_0 t)\hat S_{x/y} + \sin(\Omega_0 t)\hat S_{y/x}$. We call this the right-handed (RH) field. When $\mathrm{sign}_y=-1$, the circular polarisation must be of the form  $\cos(\Omega_0 t)\hat S_{x/y} - \sin(\Omega_0 t)\hat S_{y/x}$, which we call the left-handed (LH) field.  Table \ref{tableclockwisecounterclockwise} shows the form that  matrices $(\hat S'_{x/y})^{eg}$ take for both regimes and the polarisation of the magnetic field required to achieve resonance.

\begin{widetext}

\begin{table}[!htb]
\centering
\begin{tabular}{c|c|c}
 & right-handed & left-handed\\\hline
 $\mathrm{sign}_y$ & $\langle e|\hat S_z + \hat I_z |e\rangle -\langle g |\hat S_z + \hat I_z|g \rangle=1$ & $\langle e|\hat S_z + \hat I_z |e\rangle -\langle g |\hat S_z + \hat I_z|g \rangle=-1$ \\\hline
$(\hat S_x')^{eg}$ & $\frac{1}{2}\begin{pmatrix}0 & |e\rangle\langle e|\hat S_+|g\rangle\langle g| \\
|g\rangle\langle g|\hat S_-|e\rangle\langle e| & 0 \\
\end{pmatrix} = \frac{\eta}{2}\hat \sigma_x$ & $\frac{1}{2}\begin{pmatrix}0 & |e\rangle\langle e|\hat S_-|g\rangle\langle g| \\
|g\rangle\langle g|\hat S_+|e\rangle\langle e| & 0 \\
\end{pmatrix} = \frac{\eta}{2}\hat \sigma_x$ \\\hline
$(\hat S_y')^{eg}$ & $\frac{i}{2}\begin{pmatrix}0 & -|e\rangle\langle e|\hat S_+|g\rangle\langle g| \\
|g\rangle\langle g|\hat S_-|e\rangle\langle e| & 0 \\
\end{pmatrix} = \frac{\eta}{2}\hat \sigma_y$ & $\frac{i}{2}\begin{pmatrix}0 & |e\rangle\langle e|\hat S_-|g\rangle\langle g| \\
-|g\rangle\langle g|\hat S_+|e\rangle\langle e| & 0 \\
\end{pmatrix} =- \frac{\eta}{2}\hat \sigma_y$ \\\hline
$(\hat H_o^{\varsigma})^{eg}$ & $\frac{\Omega_0}{2}\hat \sigma_z$ & $\frac{\Omega_0}{2}\hat \sigma_z$ \\\hline 
rotating field & $\cos(\Omega_0 t)\hat S_{x/y} +\ sin(\Omega_0 t)\hat S_{y/x}$ & $\cos(\Omega_0 t)\hat S_{x/y} - \sin(\Omega_0 t)\hat S_{y/x}$ \\\hline
\end{tabular}
\caption{Pauli operators for the truncated 2-level system under resonance in the right-handed and left-handed regimes. The bottom row indicates the polarisation that the $V_{x/y}(t)$ term must have in order to effect a $e^{\pm i\frac{\theta}{2}\hat \sigma_{x/y}}$ operator in the rotating frame.  }\label{tableclockwisecounterclockwise}
\end{table}

\end{widetext}

\newpage

\section{Master equation derivation}\label{masterequationderivation}

Taking the Hamiltonian from Eq.\eqref{HAM} and adding to it a perturbative term involving independent temporal magnetic-field fluctuations in all three spatial dimensions, all of which take a Gaussian distribution with mean 0 and variance $\alpha_n^2$, gives in the interaction picture:

\begin{equation}
\tilde H(t) = \sum_{n=1}^3\omega_n(t)\tilde S_n(t) +\omega_n(t)\delta\tilde I_n(t).
\end{equation}     

\noindent where $\{S_n\}$ and $\{I_n\}$ are the $n$-axis electron and nuclear spin operators respectively, and $\omega(t)$ is the electron Zeeman frequency at time $t$. As before, $\delta$ represents the ratio of the nuclear to electronic Zeeman frequencies, and as it is small we may  ignore the nuclear term. We then  write the Liouville-von Neumann equation in differential-integral form and take the average over the field fluctuations. Noting that $\langle \frac{d}{dt}\tilde\rho(t)\rangle=\frac{d}{dt}\langle \tilde\rho(t)\rangle$ we may write this as:

\begin{eqnarray}
&&\frac{d}{dt}\langle\tilde\rho(t)\rangle=i\sum_{n=1}^3\langle\omega_n(t)\rangle\left[\langle\tilde\rho(t)\rangle,\tilde S_n(t) \right]- \nonumber \\ &&\sum_{n=1}^3 \int_{t_0}^tds\langle\omega_n(t)\omega_n(s)\rangle \left[\left[\langle\tilde\rho(t)\rangle,\tilde S^\dagger_n(t) \right] ,\tilde S_n(s)\right]
\end{eqnarray}

\noindent where assigning  $\tilde S_n(t) = \tilde S_n^\dagger(t)$ is  valid as it is a Hermitian operator. $\langle\tilde\rho(t)\rangle$ is the  density operator for the nuclear-electronic system, averaged either over an ensemble of such systems, or over many repeated experiments on the same system.   Here, we have assumed that the field fluctuation statistics are independent of the quantum state of our system.  These assumptions and approximations lead to a Born-Markov master equation.  As $\omega_n(t)$ follows a Gaussian distribution with mean 0, the first term of this equation vanishes. Given that the correlation function drops to zero at finite values, we may set the integration limits to $t_0=0$ and $t = \infty$, and  change the integration constant to $\tau = t-s$. As the temporal fluctuation takes a Gaussian distribution, we may set our correlation functions to be another Gaussian function of the form
\begin{equation}
 \langle\omega_{n}(t)\omega_{n}(t - \tau)\rangle=\frac{\alpha_{n}^2}{2\sqrt{\pi \chi_n}}e^{-\frac{\tau^2}{4 \chi_n}}
\end{equation} 

\noindent where $\chi_n = \left(\frac{d B_n}{d t}\right)^{-1}$ is an inverse function of how fast the magnetic field fluctuates. In the limiting case of $\frac{d B_n}{d t}\to\infty$, the correlation function tends to $\alpha_n^2\delta(\tau)$. To aid our calculation, we may write the noise operators in the basis of the eigenstates of $\hat H_0$ labeled $\{|a\rangle$,$|b\rangle\}$:

\begin{eqnarray}
\hat  S_n &=& \sum_\Omega \hat  S_n(\Omega)  \nonumber \\ \hat  S_n(\Omega)&=& \sum_{a,b}\delta(\omega_{ba}-\Omega)|a\rangle\langle a|\hat  S_n|b\rangle\langle b |.
\end{eqnarray}

\noindent Moving such operators  to the interaction picture simply gives $ e^{-i\Omega t } \hat  S_n(\Omega)$.  This gives

\begin{eqnarray}
\frac{d}{dt}\langle\rho(t)\rangle&=&-\alpha_n^2\sum_{n=1}^3\sum_{\Omega,\Omega'} \int_{0}^\infty \frac{d\tau}{2\sqrt{\pi \chi_n}}e^{-\frac{\tau^2}{4\chi_n}}e^{i\Omega \tau}e^{it(\Omega'-\Omega)} \times \nonumber \\
&&\left[\left[\langle\tilde\rho(t)\rangle,\hat  S^\dagger_n(\Omega') \right] ,\hat  S_n(\Omega)\right].
\end{eqnarray}  

 \noindent If $\alpha^2 \ll \Omega,\Omega'$, meaning that the dynamic time scale of our system is much shorter than that of the decoherence caused by the magnetic-field fluctuation, which is a reasonable assumption  for systems of interest, we may make the rotating wave  approximation (often also referred to as the secular approximation) and drop all terms where $\Omega \neq \Omega'$. Furthermore, noting that $[A,[B,C]]=ABC - BCA + H.C.$ (Hermitian conjugate) if $A,B,C$ are Hermitian operators leads to  :

\begin{eqnarray}
\frac{d}{dt}\langle\rho(t)\rangle&=&\alpha_n^2\sum_{i=n}^3\sum_{\Omega} \int_{0}^\infty \frac{d\tau}{2\sqrt{\pi \chi_n}}e^{-\frac{\tau^2}{4\chi_n}}e^{i \Omega\tau} \times \nonumber \\
&&\left( \hat  S^\dagger_n(\Omega)\langle\tilde\rho(t)\rangle\hat  S_n(\Omega)- \langle\tilde\rho(t)\rangle\hat  S^\dagger_n(\Omega) \hat S_n(\Omega) \right) \nonumber \\
&& \ + \mathrm{H. c.}\label{masterequationbefore}
\end{eqnarray}   

\noindent We may  decompose the integrand to 

\begin{equation}
\int_0^\infty \frac{d\tau}{2\sqrt{\pi\chi_n}}e^{-\frac{\tau^2}{4\chi_n}}e^{i \Omega\tau}=\upsilon_n(\Omega) + i\Upsilon_n(\Omega)
\end{equation}

\noindent where:

\begin{eqnarray}
\upsilon_n(\Omega) &=& \frac{1}{2}  \int_{-\infty}^{\infty}\frac{d\tau}{2\sqrt{\pi\chi_n}}e^{-\frac{\tau^2}{4\chi_n}}e^{i \Omega\tau}=\frac{1}{2}e^{-\chi_n\Omega^2} \nonumber \\
\Upsilon_n(\Omega) &=&\frac{1}{2i}\int_0^{\infty} \frac{d\tau}{2\sqrt{\pi \chi_n}}e^{-\frac{\tau^2}{4\chi_n}}\left( e^{i \Omega\tau} - e^{-i \Omega\tau} \right).
\end{eqnarray}

\noindent Plugging these expressions into Eq.\eqref{masterequationbefore}, and moving back into the Schrodinger picture, gives us our final master equation:

\begin{eqnarray}
&&\frac{d}{dt}\langle\rho(t)\rangle = i\left[\langle\rho(t)\rangle,\hat H_0 + \hat H_{LS} \right]\nonumber  +\alpha_n^2\sum_{n=1}^3\sum_{\Omega} e^{-\chi_n \Omega^2} \times \nonumber \\
&&\left( \hat  S^\dagger_n(\Omega)\langle\tilde\rho(t)\rangle\hat  S_n(\Omega)-\frac{1}{2}\left[ \langle\tilde\rho(t)\rangle,\hat  S^\dagger_n(\Omega) \hat S_n(\Omega)\right]_+  \right)
 \end{eqnarray}

\noindent where

\begin{equation}
H_{LS}=\sum_n\sum_\Omega\gamma_n(\Omega)\hat S_n^\dagger(\Omega)\hat S_n(\Omega)
\end{equation}

\noindent is the Lamb shift and changes the energy levels of the system. This is a negligible effect and hence can be ignored. We use $\left[A,B \right]_+:= AB+BA$ as the anticomutator operator.

\end{document}